\documentclass[twocolumn,showpacs,showkeys,nofootinbib,superscriptaddress]{revtex4}
\usepackage{graphicx}
\usepackage{amsmath,amsfonts,amssymb,bm}

\begin{document}

\title{Experimental study of $\mu$--atomic and $\mu$--molecular
processes in pure helium and deuterium--helium mixtures}

\author{V.M.~Bystritsky}
\altaffiliation{Corresponding author}
\email{bystvm@nusun.jinr.ru}
\affiliation{Joint Institute for Nuclear Research, Dubna 141980,
Russia}

\author{V.F.~Boreiko}
\affiliation{Joint Institute for Nuclear Research, Dubna 141980,
Russia}

\author{W.~Czapli\'nski}
\affiliation{University of Mining and Metallurgy,
Fac.~Phys.~Nucl.~Techniques, PL--30059 Cracow, Poland}

\author{M.~Filipowicz}
\affiliation{University of Mining and Metallurgy, Fac.~of Fuels and
Energy, PL--30059 Cracow, Poland}

\author{V.V.~Gerasimov}
\affiliation{Joint Institute for Nuclear Research, Dubna 141980,
Russia}

\author{O.~Huot}
\affiliation{Department of Physics, University of Fribourg, CH--1700
Fribourg, Switzerland}

\author{P.E.~Knowles} 
\affiliation{Department of Physics, University of Fribourg, CH--1700
Fribourg, Switzerland}

\author{F.~Mulhauser}
\altaffiliation[Present address: ]{University of Illinois at
  Urbana--Champaign, Urbana, Illinois 61801, USA}
\affiliation{Department of Physics, University of Fribourg, CH--1700
Fribourg, Switzerland}

\author{V.N.~Pavlov}
\affiliation{Joint Institute for Nuclear Research, Dubna 141980,
Russia}

\author{N.P.~Popov}
\altaffiliation{Visiting Professor}
\affiliation{University of Mining and Metallurgy,
Fac.~Phys.~Nucl.~Techniques, PL--30059 Cracow, Poland}

\author{L.A.~Schaller}
\affiliation{Department of Physics, University of Fribourg, CH--1700
Fribourg, Switzerland}

\author{H.~Schneuwly}
\affiliation{Department of Physics, University of Fribourg, CH--1700
Fribourg, Switzerland}

\author{V.G.~Sandukovsky}
\affiliation{Joint Institute for Nuclear Research, Dubna 141980,
Russia}
\author{V.A.~Stolupin}
\affiliation{Joint Institute for Nuclear Research, Dubna 141980,
Russia}

\author{V.P.~Volnykh} 
\affiliation{Joint Institute for Nuclear Research, Dubna 141980,
Russia}

\author{J.~Wo\'zniak}
\affiliation{University of Mining and Metallurgy,
Fac.~Phys.~Nucl.~Techniques, PL--30059 Cracow, Poland}

\date{\today}

\begin{abstract}
We present experimental results of $\mu$--atomic and
$\mu$--molecular processes induced by negative muons in pure helium
and helium--deuterium mixtures.
The experiment was performed at the Paul Scherrer Institute
(Switzerland).
We measured muonic x--ray $K$ series transitions relative intensities
in $(\mu{}^{3,4}\mathrm{He})^*$ atoms in pure helium as well as in
helium--deuterium mixture.
The muon stopping powers ratio between helium and deuterium atoms and the
$d \mu{}^3 \mathrm{He}$ radiative decay probability of for two
different helium densities in $\mathrm{D}_2 + {}^3\mathrm{He}$ mixture
were also determined.
Finally, the $\mathrm{q}_{1s}^{\mathrm{He}}$ probability for a $d\mu$
atom formed in an excited state to reach the ground state was measured
and compared with theoretical calculations using a simple cascade
model.
\end{abstract}


\maketitle


\section{Introduction}
\label{sec:introduction}

The experimental study of atomic and molecular processes induced by
negative muons captured in hydrogen and helium provides a test of
many--body calculations~\cite{petit01} comprising different methods of
atomic, molecular, and nuclear physics.
In spite of about 50~years of
experimental~\cite{bystr95d,bystr95c,marsh01,nagam01,petit92} and
theoretical~\cite{fesen96,faifm01,cohen99a,marku01,adamc01} studies
for processes occurring in helium and deuterium, as well as
helium--deuterium mixtures, there exist still some open questions.
The most important are listed here: 
\begin{itemize}
  \item[--] direct atomic muon capture in $h$--He mixtures ($h
            = \mathrm{H}_2, \mathrm{D}_2, \mathrm{T}_2$ and
            $\mathrm{He} = {}^{3}\mathrm{He},
            {}^{4}\mathrm{He}$);
  \item[--] initial population of $\mu h$ and $\mu\mathrm{He}$ excited
            states for various deexcitation processes of muonic atoms
            (e.g., Stark mixing, Auger and Coulomb deexcitation
            processes~\cite{marku94,bystr96,jense01,popov01,bystr00});
  \item[--] muon transfer between excited states of $\mu h$ and
            $\mu\mathrm{He}$~\cite{bystr99,bystr00,kravt94,sakam99,augsb03};
  \item[--] the probability $\mathrm{q}_{1s}$ to reach the $\mu h$
            ground state in a $h$--He
            mixture~\cite{kottm93,bystr99,bystr00,augsb03,tresc98c,lauss96};
  \item[--] ground state muon transfer from $\mu h$ to helium via 
            the intermediate $2p\sigma$ molecular state,
            $h\mu\mathrm{He}$~\cite{bystr83b,arist81,tresc98,kravt93,gartn00},
            and the subsequent decay to the unbound $1s\sigma$
            state~\cite{bystr95d,augsb03,tresc98c,kinox93,czapl97b}.
\end{itemize}

In the case of a deuterium--helium mixture the $(d\mu \mathrm{He})^*$
molecule, created in $d\mu + \mathrm{He}$ collisions, has three
possible decay channels:
\newcounter{bean}
\setcounter{bean}{0}
\renewcommand{\theequation}{1\alph{bean}}
\begin{eqnarray} 
\label{eq1}  
d \mu + {\mathrm{He}} \stackrel{\lambda_{d \mu
\mathrm{He}}}{\longrightarrow} [(d \mu {\mathrm{He}} )^* e ] &
\stackrel{\lambda _{\gamma}}{\rightarrow} & [(\mu
{\mathrm{He}})^+_{1s} e] + d +\gamma \addtocounter{bean}{1} \\
& \stackrel{\lambda_p}{\rightarrow} & [(\mu {\mathrm{He}})^+_{1s} e] +
d \addtocounter{bean}{1} \\
& \stackrel{\lambda_e}{\rightarrow} & (\mu {\mathrm{He}})^+_{1s} + d +
e \, .  \addtocounter{bean}{1}
\end{eqnarray}
\renewcommand{\theequation}{\arabic{equation}}
\addtocounter{equation}{-2}
Here, $\lambda _{\gamma}$ is the $(d\mu \mathrm{He})^*$ molecular
decay channel for the 6.85~keV $\gamma$--ray emission, $\lambda_e$ for
the Auger decay, and $\lambda_p$ for the break--up process.
The $(d\mu \mathrm{He})$ molecule is formed, with a rate $\lambda_{d
\mathrm{He}}$, in either a $J = 0$ or a $J=1$ rotational state ($J$
denotes the total angular momentum of the three particles).
The $J = 1$ state is mostly populated at slow $d\mu$--He collisions.
The $J=1 \to J=0$ deexcitation due to inner or outer Auger transition
is also possible~\cite{czapl96c,czapl02,czapl02a}.
In principle it competes with the decay processes of Eq.~(\ref{eq1}),
and can be followed by another decay due to nuclear deuterium--helium
fusion from the $J = 0$ state~\cite{bogda98,penko97}.

In this paper we present experimental results for fundamental
characteristics of $\mu$--atomic (MA) and $\mu$--molecular (MM)
processes in a $\mathrm{D}_2 +{}^{3}\mathrm{He}$ mixture, namely the
muon stopping power ratio, the $\mathrm{q}_{1s}^{\mathrm{He}}$
probability, the radiative branching ratio for the radiative decay of the
$(d\mu {}^{3}\mathrm{He})^*$ molecule (\ref{eq1}a), and delayed Lyman
series transitions in $\mu\mathrm{He}$ atoms for two different target
densities and at nearly constant helium concentrations.
Results for relative intensities of $\mu\mathrm{He} \,K$ series
transitions in pure ${}^{3,4}\mathrm{He}$ and $\mathrm{D}_2 +
{}^{3}\mathrm{He}$ for different target densities are also presented.


\section{Experimental conditions}
\label{sec:experiment}

A study of MA and MM processes mentioned above requires the
simultaneous use of miscellaneous detectors appropriate for the
detection of the muon beam, the muonic x~rays of $\mu h$ and
$\mu\mathrm{He}$ atoms (formed in the target due to direct muon
capture by the correspondent nuclei or due to muon transfer from
hydrogen to helium), products of nuclear reactions occurring in $\mu
h-\mathrm{He}$ complexes, and muon decay electrons.
Detection of the latter is necessary not only for yield normalization
but also for background reduction.
This was realized by requesting that the muon survives
atomic and molecular processes. 
Thus, muon decay electrons were detected within a certain time
interval after the principal particle detection.
For a precise measurement of the characteristics of MA and MM
processes the detection system and the associated electronics should
posses high energy and time resolutions.

\begin{figure}[ht]
  \centerline{
\setlength{\unitlength}{0.1mm}
\begin{picture}(800,800)(0,0)
\put(390,450){\oval(200,200)}
\put(380,440){T}
\put(250,300){\line(1,0){300}}
\put(250,600){\line(1,0){300}}
\put(550,600){\line(0,-1){300}}
\put(250,600){\line(0,-1){120}}
\put(250,300){\line(0,1){120}}
\put(250,420){\line(1,0){40}}
\put(250,480){\line(1,0){40}}
\put(290,480){\line(0,-1){60}}
\put(260,520){V}
\put(130,425){\framebox(150,50)}
\put(150,438){\small Ge$_S$}
\put(300,565){\framebox(200,15)}
\put(300,570){\line(1,0){200}}
\put(130,680){\small Si$_{UP}$}
\put(220,670){\vector(1,-1){90}}
\put(300,320){\framebox(200,15)}
\put(300,330){\line(1,0){200}}
\put(130,220){\small Si$_{DO}$}
\put(220,240){\vector(1,1){80}}
\put(515,350){\framebox(15,200)}
\put(520,350){\line(0,1){200}}
\put(630,630){\small Si$_{RI}$}
\put(620,620){\vector(-1,-1){90}}
\put(220,625){\framebox(350,15)}
\put(630,715){\small E$_{UP}$}
\put(620,705){\vector(-1,-1){60}}
\put(580,290){\framebox(15,320)}
\put(670,205){\small E$_{RI}$}
\put(660,225){\vector(-1,1){60}}
\put(220,260){\framebox(350,15)}
\put(600,150){\small E$_{DO}$}
\put(620,190){\vector(-1,1){60}}
\put(220,290){\framebox(15,120)}
\put(220,490){\framebox(15,120)}
\put(100,565){\small E$_{LE}$}
\put(180,550){\vector(1,-1){40}}
\put(250,660){\line(1,0){300}}
\put(250,660){\line(0,1){100}}
\put(550,660){\line(0,1){100}}
\put(355,700){\small NE213}
\put(610,300){\line(1,0){100}}
\put(610,600){\line(1,0){100}}
\put(610,600){\line(0,-1){300}}
\put(630,435){\small Ge$_M$}
\put(250,240){\line(1,0){300}}
\put(250,240){\line(0,-1){100}}
\put(550,240){\line(0,-1){100}}
\put(380,180){\small Ge$_B$}
\put(250,80){\line(1,0){300}}
\put(250,75){\line(0,1){10}}
\put(240,30){0}
\put(400,75){\line(0,1){10}}
\put(550,75){\line(0,1){10}}
\put(520,30){10 cm}
\end{picture}
}
   \caption{Scheme of the experimental setup. The view is that of the
           incoming muon.}
   \label{fig:setup}
\end{figure}
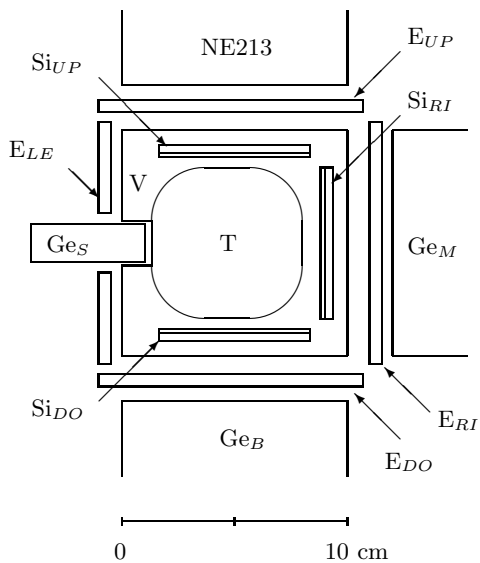

The experiment was performed at the Paul Scherrer Institute (PSI) at
the $\mu$E4 muon channel. 
It is described in details in Refs.~\cite{borei98,knowl01,bystr03}.
A schematic muon eyes view of the setup is given in
Fig.~\ref{fig:setup}.

The experimental setup was designed and developed to study nuclear
reactions in charge asymmetric muonic molecules such as $(d\mu
{}^{3}\mathrm{He})$~\cite{knowl01,maevx99,czapl96,czapl98,penko97,
bogda99,bogda98,bystr99c,bystr99d}:
\begin{equation}
  \label{eq2}
  d\mu ^3{\mathrm{He}} \to \alpha\, (3.66\, \mathrm{MeV}) + \mu + p\,
  (14.64\, \mathrm{MeV}).
\end{equation}
Charged muon--capture products were detected by three silicon
telescopes located directly in front of the kapton windows but still
within the cooled vacuum environment (Si$_{UP}$, Si$_{RI}$, and
Si$_{DO}$).
Muon decay electrons were detected by four pairs of plastic
scintillator counters (E$_{LE}$, E$_{UP}$, E$_{RI}$, E$_{DO}$) placed
around the target.
The cryogenic target body was made of pure aluminium and had different
kapton windows in order to detect in particular
\begin{itemize}
\item[--] the $\sim34$~MeV/c momentum muon beam,
\item[--] the 6.85~keV $\gamma$~rays emitted via the radiative decay
          given in Eq.~(\ref{eq1}a),
\item[--] the x--ray Lyman series transitions from the
          $\mu\mathrm{He}$ deexcitation ($K{\alpha}$ at 8.2~keV,
          $K{\beta}$ at 9.6~keV, and $K{\gamma }$ at 10.2~keV).
\end{itemize}
The 0.17~cm$^3$ germanium detector (Ge$_S$) used for the $\gamma$ and
x--ray detection was placed just behind a 55~$\mu$m thick kapton
window.

The experiment includes four groups of measurements as depicted in
Table~\ref{tab:exp-cond}.
The first two groups, I and II, are ${}^{3}\mathrm{He}$ and
${}^{4}\mathrm{He}$ measurements at different temperatures and
pressures.
The remaining measurements, III and IV, were performed with
$\mathrm{D}_2 +{}^{3}\mathrm{He}$ mixtures at two different densities.
The density $\varphi$ is normalized to the liquid hydrogen density
(LHD), $N_0 = 4.25 \times 10^{22}\, \mbox{cm}^{-3}$.
Run III was by far the longest run because its original purpose was to
measure the fusion rate in the $d\mu {}^{ 3} \mathrm{He}$ molecule,
Eq.~(\ref{eq2}), and the muon transfer rate $\lambda_{d
{}^3\mathrm{He}}$ from $d\mu$ atoms to ${}^{3}\mathrm{He}$
nuclei~\cite{knowl01}.
The germanium detector energy calibration was carried out during the
data taking period using standard sources, namely $^{60}$Co,
$^{57}$Co, $^{55}$Fe, and $^{137}$Cs.

\begin{table}[t]
\begin{ruledtabular}
       \caption{Experimental conditions, such as temperature,
               pressure, density, and helium concentration. The last
               column presents the number of muon stops in the gas.}
\label{tab:exp-cond}
\begin{tabular}{ccccccc}
Run & Gas & Temp. & Pressure & $\varphi$ & $\mathrm{c}_{{\mathrm{He}} }$ &
$N_\mathrm{stop}$ \\
&& [K] & [atm] & [LHD] & [\%] & [10$^{6}$] \\ \hline
I  &${}^{3}\mathrm{He}$ & 32.9 &--- & --- & 100 & --- \\
Ia && & 6.92 & 0.0363 && 640.4 \\
Ib && & 6.85 & 0.0359 && 338.1 \\
Ic && & 6.78 & 0.0355 && 375.3 \\
Id && & 6.43 & 0.0337 && 201.7 \\ \hline
II  &${}^{4}\mathrm{He}$ &--- & --- & --- & 100 & --- \\
IIa && 20.3 & 12.55 & 0.1060 && 239.4 \\
IIb && 19.8 & 9.69 & 0.844 && 554.1 \\
IIc && 20.0 & 4.52 & 0.039 && 32.3 \\ \hline
    & $\mathrm{D}_2 +{}^{3}\mathrm{He}$ & 32.8  & ---  & ---    & 4.96 &
  --- \\
III & & & 5.11 & 0.0585 & & 4215.6 \\ 
IV & & & 12.08 & 0.1680 & & 2615.4 \\ 
\end{tabular}
\end{ruledtabular}
\end{table}
                                                      

\section{Method of the measurement}
\label{sec:method-measurement}

\begin{figure*}[ht]
\centerline{
\setlength{\unitlength}{0.75mm}
\thicklines
\begin{picture}(200,145)(0,-10)
\put(10,85){\oval(10,10)}
\put(10,85){\makebox(0,0){$\mu^{-}$}}
\put(15,85){\vector(1,0){8}}
\put(20,90){\makebox(0,0){$W_\mathrm{D}$}}
\put(12,80){\vector(1,-1){15}}
\put(13,70){\makebox(0,0){$W_{\mathrm{He}}$}}
\put(35,90){\vector(0,1){15}}
\put(30,97){\makebox(0,0){$\mathrm{q}_{1s}$}}
\put(35,80){\vector(0,-1){15}}
\put(46,73){\makebox(0,0){$(1-\mathrm{q}_{1s})$}}
\put(33,85){\oval(20,10)}
\put(35,85){\makebox(0,0){$(\mu d )^{*}$}}
\put(33,110){\oval(20,10)}
\put(33,110){\makebox(0,0){$(\mu d)_{1s,F}$}}
\put(33,60){\oval(20,10)}
\put(33,60){\makebox(0,0){$\mu{}^3\mathrm{He}$}}
\put(35,55){\vector(0,-1){10}}
\put(28,50){\makebox(0,0){$\lambda_\mathrm{cap}^{\mathrm{He}}$}}
\put(6,22){\framebox(24,10){}}
\put(18,27){\makebox(0,0){$p + 2 n + \nu_\mu$}}
\put(40,22){\framebox(24,10){}}
\put(52,27){\makebox(0,0){$d + n + \nu_\mu$}}
\put(25,0){\framebox(20,10){}}
\put(35,5){\makebox(0,0){$t + \nu_\mu$}}
\put(35,45){\vector(1,-1){13}}
\put(48,40){\makebox(0,0){$\lambda_\mathrm{cap}^{d}$}}
\put(35,45){\vector(-1,-1){13}}
\put(22,40){\makebox(0,0){$\lambda_\mathrm{cap}^{p}$}}
\put(35,45){\vector(0,-1){35}}
\put(42,16){\makebox(0,0){$\lambda_\mathrm{cap}^{t}$}}
\put(43,110){\line(1,0){7}}
\put(58,110){\makebox(0,0){$\varphi \, \mathrm {\mathrm c}_{\mathrm d}
\lambda_{dd\mu}$}}
\put(66,110){\vector(1,0){7}}
\put(40,105){\line(2,-3){28}} 
\put(70,60){\makebox(0,0){$\varphi \,
\mathrm {\mathrm c}_{{}^3\mathrm{He}} \lambda_{d{}^3\mathrm{He}}$}} 
\put(72,57){\vector(2,-3){8}}
\put(75,15){\framebox(35,30){}}
\put(92,35){\makebox(0,0){$[d\mu{}^3\mathrm{He}]^+ + e$}}
\put(85,25){\makebox(0,0){2p$\sigma$}}
\put(100,25){\makebox(0,0){J=1}}
\put(150,15){\framebox(35,30){}}
\put(167,35){\makebox(0,0){$[d\mu{}^3\mathrm{He}]^+ + e$}}
\put(160,25){\makebox(0,0){2p$\sigma$}}
\put(175,25){\makebox(0,0){J=0}}
\put(110,30){\line(1,0){15}}
\put(132,30){\makebox(0,0){$\tilde{\lambda}_{10}$}}
\put(138,30){\vector(1,0){12}}
\put(100,60){\framebox(70,30){}}
\put(135,80){\makebox(0,0){$\alpha + \mu ^{-} + p (14.7\:\mathrm{MeV})$}}
\put(135,70){\makebox(0,0){$\mu{}^{5}\!\mathrm{Li}+\gamma(16.4\:\mathrm{MeV})$}}
\put(115,0){\framebox(30,16){}}
\put(130,8){\makebox(0,0){$[\mu {}^3\mathrm{He} + d]$}}
\put(90,15){\line(0,-1){5}}
\put(90,10){\vector(1,0){25}}
\put(170,15){\line(0,-1){5}}
\put(170,10){\vector(-1,0){25}}
\put(80,2){\makebox(0,0){$\lambda^{J=1}_{p}$,}}
\put(93,2){\makebox(0,0){$\lambda^{J=1}_{e}$,}}
\put(107,2){\makebox(0,0){$\lambda^{J=1}_{\gamma}$}}
\put(153,2){\makebox(0,0){$\lambda^{J=0}_{p}$,}}
\put(166,2){\makebox(0,0){$\lambda^{J=0}_{e}$,}}
\put(179,2){\makebox(0,0){$\lambda^{J=0}_{\gamma}$}}
\put(90,45){\line(0,1){25}}
\put(90,70){\vector(1,0){10}}
\put(180,45){\line(0,1){25}}
\put(180,70){\vector(-1,0){10}}
\put(93,75){\makebox(0,0){$\lambda^{J=1}_f$}}
\put(180,75){\makebox(0,0){$\lambda^{J=0}_f$}}
\put(83,110){\oval(20,10)}
\put(83,110){\makebox(0,0){$dd \mu  $}}
\put(93,110){\vector(1,0){17}}
\put(110,100){\line(0,1){22}}
\put(110,122){\vector(1,0){15}}
\put(117,125){\makebox(0,0){$\beta_F$}}
\put(125,116){\line(0,1){12}}
\put(125,128){\vector(1,0){15}}
\put(133,125){\makebox(0,0){$\omega_{d}$}}
\put(150,128){\oval(20,10)}
\put(150,128){\makebox(0,0){$\mu {}^3\mathrm{He}$}}
\put(165,128){\makebox(0,0){$+ n$}}
\put(125,116){\vector(1,0){15}}
\put(131,119){\makebox(0,0){$1\!-\!\omega_{d}$}}
\put(170,116){\makebox(0,0){$\:{}^3\mathrm{He} + \mu^{-} + n {\mathrm{(2.45~MeV)}}$}}
\put(110,100){\vector(1,0){30}}
\put(120,103){\makebox(0,0){$1-\beta_F$}}
\put(168,100){\makebox(0,0){$t + \mu^{-} + p\:{\rm (3.02~MeV)}$}}
\put(150,133){\line(0,1){2}}
\put(150,135){\line(-1,0){150}}
\put(0,135){\vector(0,-1){40}}
\put(0,95){\line(0,-1){35}}
\put(0,60){\vector(1,0){23}}
\put(130,0){\line(0,-1){3}}
\put(130,-3){\line(-1,0){130}}
\put(0,37){\line(0,1){23}}
\put(0,-3){\vector(0,1){40}}
\end{picture}
}
      \caption{Scheme of $\mu$--atomic and $\mu$--molecular processes
               in the $\mathrm{D}_2 + {}^{3}\mathrm{He}$ mixture.
               Details about all processes and rates are found in
               Ref.~\cite{knowl01}.}
      \label{fig:2}
\end{figure*}
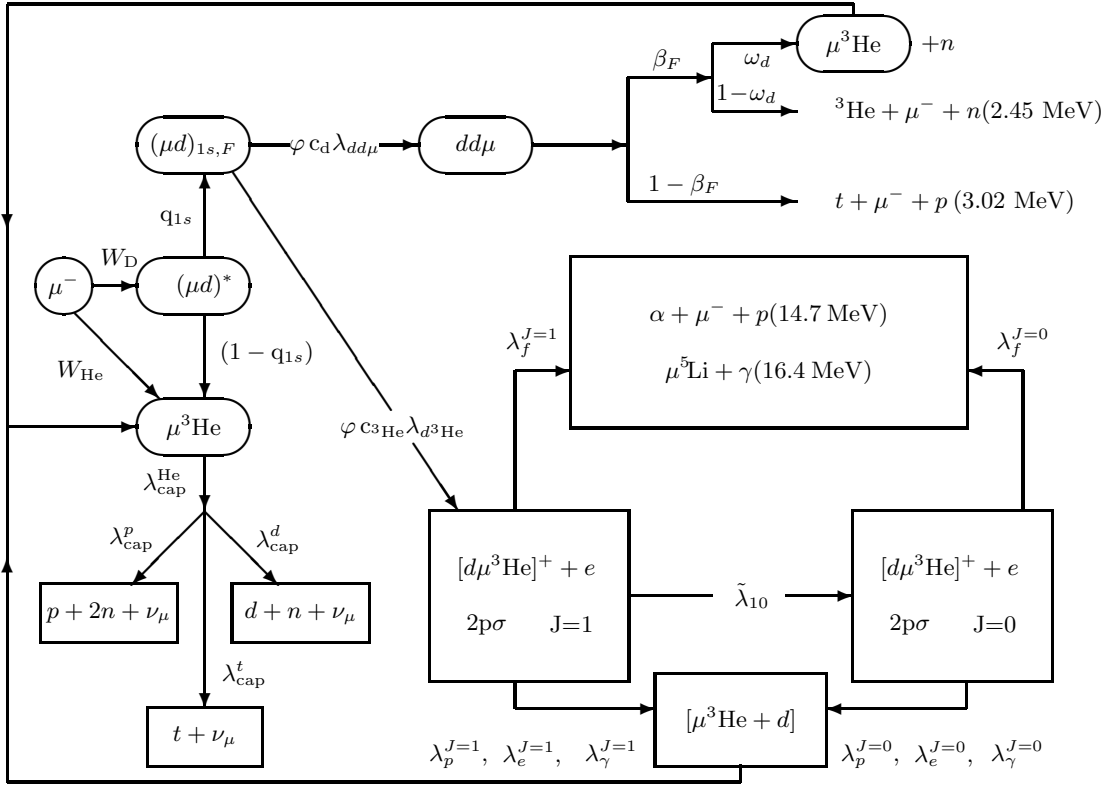

The MA and MM processes occurring in $\mathrm{D}_2
+{}^{3}\mathrm{He}$ mixtures after the muon has stopped in a target volume is
explained in detail in Ref~\cite{bystr03}.
The main characteristics are shown in Fig.~\ref{fig:2}.
Depending on the time elapsed after a muon stop in a mixture one can
distinguish between prompt and delayed MA processes.

The following processes are considered as prompt ones: 
\begin{enumerate}
\item[--] the slowing down of muons entering a target to velocities
          enabling an atomic capture into excited states of $\mu h$ or
          $\mu\mathrm{He}$, with a characteristic moderation time
          $t_{mod} < 10^{-9}$~s for target densities $\varphi >
          10^{-3}$~\cite{fesen96,schmi98,hause93b,adamc96b,kottm99};
\item[--] the formation of excited muonic atoms, $(\mu h)^*$ ,
          $(\mu\mathrm{He})^*$, $t_{form} \sim
          10^{-11}$~s~\cite{jense01};
\item[--] the cascade transitions in $(\mu h)^*$ and
          $(\mu\mathrm{He})^*$ muonic atoms, $t_{casc} \sim
          10^{-11}$~s~\cite{koike01};
\item[--] the muon transfer from exited states of $(\mu h)^*$ to
          helium (occurring in $\mathrm{D}_2 +{}^{3}\mathrm{He}$
          mixtures), $t \leq
          10^{-10}$~s~\cite{bystr99,sakam99,augsb03,tresc98c}.
\end{enumerate}
The delayed processes are: 
\begin{enumerate}
\item[--] the ground state muon transfer from muonic deuterium to
          helium~\cite{tresc98c,gartn00};
\item[--] the formation of excited $(d\mu {}^{3}\mathrm{He})^*$
          molecules (with the subsequent prompt decay after about
          $10^{-11}$~s~\cite{kinox93,czapl97b}).
\end{enumerate}


\subsection{Pure helium}
\label{su:pure}

One of the main characteristics of MA processes occurring in pure
helium are absolute and relative intensities of muonic $K$ series
x--ray transitions in $(\mu\mathrm{He})^*$ atoms.
Their knowledge provides important information about the excited
states initial population of the $\mu\mathrm{He}$ atoms and the
dynamics of deexcitation.
According to the above given classification of MA processes and the
conditions of runs I and II it is clear that only prompt $K$
series transitions from $\mu\mathrm{He}$ were observed.
Events detected by the germanium detector within a time range $-30
\,\mathrm{ns} \leq t_\gamma \leq 30 \, \mathrm{ns}$ around time $t=0$
(defined as the muon stop time) were classified as prompt.
The chosen time range is a consequence of the detector and its related
electronic time resolution.
The relative intensities, $I^{\mathrm{He}}_x$, of the $Kx$~lines ($x \equiv
\alpha, \beta, \gamma$) are:
\begin{equation}
  \label{eq4}
  I^{\mathrm{He}}_x = \frac{Y^{\mathrm{He}}_x}{Y_{tot}^{\mathrm{He}}}
  \qquad \mbox{with} \qquad \sum_{x= \alpha, \beta, \gamma}
  I^{\mathrm{He}}_x = 1\, ,
\end{equation}
where $Y^{\mathrm{He}}_\alpha, Y^{\mathrm{He}}_\beta,
Y^{\mathrm{He}}_\gamma$ are the yields of $\mu\mathrm{He}$
$Kx$~lines with energies 8.17~keV, 9.68~keV, and 10.2~keV,
respectively.
These yields are determined as follows:
\begin{equation}
  \label{eq5}
  Y^{\mathrm{He}}_x = \frac{N^{\mathrm{He}}_x}{(1-\eta_x)
  \varepsilon_x } \, , \qquad Y_{tot}^{\mathrm{He}} =
  \sum_{x= \alpha, \beta, \gamma} Y^{\mathrm{He}}_x \,
\end{equation}
with $Y_{tot}^{\mathrm{He}}$ being the total yield of all $Kx$~lines.
The quantities $N^{\mathrm{He}}_x$ are the prompt events numbers
corresponding to the $\mu\mathrm{He}$ $Kx$~lines.
$\eta_x$ are the total attenuation coefficients of the corresponding
$Kx$~lines and $\varepsilon _x$ are the corresponding detection
efficiencies.
The $I^{\mathrm{He}}_\gamma$ intensity is the cumulative photon yield
of the Lyman series $n \geq 4$.

In fact, only detection efficiency ratios $(\varepsilon_{x\alpha} =
\varepsilon_x /\varepsilon_\alpha)$ are required for the determination
of the relative intensities.
Therefore Eq.~(\ref{eq4}) can be rewritten as
\begin{equation}
\label{eq4a}
I^{\mathrm{He}}_x = \frac{N^{\mathrm{He}}_x}
{N_{tot}^{\mathrm{He}} (1-\eta_x)\varepsilon_{x\alpha}}\, , 
\end{equation}
with
\begin{equation}
\label{eq4b}
N_{tot}^{\mathrm{He}} = \sum_{x= \alpha, \beta, \gamma}
\frac{N^{\mathrm{He}}_x}{(1-\eta_x)\varepsilon_{x\alpha}}
\end{equation}
being the total yield normalized to the detection efficiency
$\varepsilon_\alpha$.
This fact significantly increases the accuracy of $I^{\mathrm{He}}_x$
measured in experiment.
The corresponding errors were mainly due to insufficient knowledge of
the respective attenuation coefficients.
One can expect, however, that attenuation coefficients, as compiled in
Ref.~\cite{storm70}, differ only slightly because the differences
between energies of $Kx$~lines ($\Delta E_{\beta - \alpha} = E(K\beta)
- E(K\alpha) = 1.51$~keV, $\Delta E_{\gamma - \alpha}\ = E(K\gamma) -
E(K\alpha) = 2.03$~keV) are relatively small.
Recent experimental results also confirmed this assumption (see
Refs.~\cite{augsb03,gartn00}).

The detection efficiencies, $\varepsilon_x$ are determined using
Eqs.~(\ref{eq4}) and~(\ref{eq5}) via
\begin{equation}
\label{eq19a}
\varepsilon_x = \frac{N^{\mathrm{He}}_x}{N_\mathrm{stop}^{\mathrm{He}}
I^{\mathrm{He}}_x}\, ,
\end{equation}
where $N_\mathrm{stop}^{\mathrm {He}}$ is the number of muons stopping
in helium, given in Table~\ref{tab:exp-cond}.
%
%
For an accurate determination of the $K$ series transitions
attenuation coefficient we performed Monte Carlo (MC) calculations
taking into account the experimental geometry and all material layers
placed between the x--ray emission and the germanium detector.
The total attenuation coefficient $\eta_x$ of each $Kx$~line includes
the x--ray attenuation while passing through the gas, the target and
chamber kapton window, and through the germanium detector Be window
taken from Ref.~\cite{hubbe82}.
We obtained $\eta_\alpha = 0.156$, $\eta_\beta = 0.085$, and
$\eta_\gamma = 0.075$.
 
A significant reduction of the germanium detector background was
achieved by using delayed coincidences between x~rays and electrons.
This method is called the ``del-$e$'' criterion.
Ground state muonic helium atoms disappear mainly by muon decay,
\begin{equation}
  \label{eq3}
  \mu ^- \to e^- + \nu_\mu + \bar{\nu}_e \, ,
\end{equation} 
and by nuclear muon capture (with proton, deuteron, or triton
emission~\cite{maevx96,phili75,bystr03}).
The average disappearance rate is
\begin{equation}
  \label{eq3a}
  \lambda_{\mathrm{He}} = \lambda_0 + \lambda_{cap}^{\mathrm{He}}
  \approx 0.457 \times 10^6 \, \mbox{s}^{-1} \, ,
\end{equation}
where $\lambda_0 = 0.455 \times 10^6 \, \mbox{s}^{-1}$ and
$\lambda_{cap}^{\mathrm{He}} = 2216(70) \,
\mbox{s}^{-1}$~\cite{maevx96}.
Thus, delayed electrons were measured during a time interval
corresponding to two $\mu\mathrm{He}$ atom life times ($\tau
_{\mathrm{He}} = 2.19 \, \mu$s~\cite{suzuk87}).

The relative intensities of $Kx$~lines, $I^{\mathrm{He}}_{x-
\mathrm{e}}$, detected in coincidences with muon decay electrons, are
given by
\begin{equation}
  \label{eq7}
  I^{\mathrm{He}}_{x- \mathrm{e}} = \frac{1}{\varepsilon_\mathrm{e}
  f_t} \frac{N^{\mathrm{He}}_{x -\mathrm{e}}}
  {N_{tot,\mathrm{e}}^{\mathrm{He}} (1-\eta_x)\varepsilon_{x\alpha}}
\end{equation}
with
\begin{equation}
  \label{eq8}
  N_{tot,\mathrm{e}}^{\mathrm{He}} = \frac{1}{\varepsilon_\mathrm{e}
  f_t} \sum_{x= \alpha, \beta, \gamma} \frac{N^{\mathrm{He}}_{x
  -\mathrm{e}}}{(1-\eta_x)\varepsilon_{x\alpha}}
\end{equation}
where $N^{\mathrm{He}}_{x - \mathrm{e}}$ are the number of events in
pure helium detected by the germanium detector in coincidence with
muon decay electrons within a fixed time interval $\Delta t = t_e -
t_\gamma$, with $t_\gamma$ and $t_e$ the time of the germanium and
decay electron counters, respectively.
Both times are measured relative to the muon stop time $t=0$.
$\varepsilon_\mathrm{e}$ is the detection efficiency of muon decay
electrons and the time factor
\begin{equation}
  \label{eq8a}
  f_t = 1 - e^{-\lambda_{\mathrm{He}} \Delta t}
\end{equation}
is the probability that a muon decays in the ground state of
$\mu\mathrm{He}$ during the time interval $\Delta t$.

It should be noted, that the coefficient $\varepsilon_\mathrm{e} f_t$
is not required as an absolute number for the determination of the
intensities $I^{\mathrm{He}}_{x - \mathrm{e}}$ as it enters the
numerator and denominator of Eq.~(\ref{eq7}) in the same manner.
However, it is needed for the $\mathrm{D}_2 +{}^{3}\mathrm{He}$
analysis.
The quantity $\varepsilon_\mathrm{e} f_t$ is determined by comparing
Eqs.~(\ref{eq4a}) and~(\ref{eq7}) yielding
\begin{equation}
\label{eq9}
\varepsilon_\mathrm{e} f_t = \frac{N^{\mathrm{He}}_{x - \mathrm{e}
}}{N^{\mathrm{He}}_{x}} \, .
\end{equation}

Another interesting problem is the study of $\mu\mathrm{He}$ atoms in
excited metastable $2s$ states.
One can expect, according to
Refs.~\cite{cohen82,vonar84,eckha86,cohen88s}, that the
$(\mu\mathrm{He})_{2s}$ atom population varies between 5\% and 7\%
under our experimental conditions for runs I and II\@.
The two possible channels of $2s \to 1s$ deexcitation are two--photon
transition with a rate $\lambda_{2\gamma} \sim 1.06 \times 10^5 \,
\mbox{s}^{-1}$~\cite{johns72,bache84} and the Stark $2s \to 2p \to 1s$
deexcitation~\cite{cohen82,vonar84,eckha86} induced by collisions of
$(\mu\mathrm{He})_{2s}$ atoms with the surrounding atoms or molecules.
The corresponding rate for the experimental conditions of runs I and II is
$\lambda \sim 2.2 \times 10^7 \, \mbox{s}^{-1}$.
If the time of Stark induced transitions is shorter than the
resolution time of the germanium detector the corresponding $K\alpha$
transition would be experimentally classified as a prompt event.
Otherwise, it would be possible to extract an upper bound for Stark
induced transition rates.


\subsection{$\mathrm{D}_2 +{}^{3}\mathrm{He}$ mixtures}
\label{su:mixture}

In a $\mathrm{D}_2 +{}^{3}\mathrm{He}$ mixture one observes $Kx$~lines
arising from the deexcitation of $\mu\mathrm{He}$ atoms formed not
only due to direct muon capture by helium nuclei (as in pure helium)
but also due to muon transfer from muonic deuterium to helium.
Because the $d\mu$ atoms deexcitation time is of the order $10^{11} \,
\mbox{s}^{-1}$ (under our experimental conditions) the corresponding
emission of $K$ series transitions occurs practically immediately
after a muon stop in the mixture and can be classified as a prompt
event.
%
%
Muons are captured by $\mathrm{D}_2$ and ${}^{3}\mathrm{He}$ according
to the capture law~\cite{bystr95d}.
Information about relative rates of atomic muon capture by deuterons
and helium nuclei in a $\mathrm{D}_2 +{}^{3}\mathrm{He}$ mixture as
well as the probability that an excited $(d\mu)^*$ atom reaches its
ground state when the muon also has the possibility of transferring
directly from an excited state to a heavier nucleus, in our case
helium ($\mathrm{q}_{1s}^{\mathrm{He}}$ probability) is of
unquestionable importance for understanding kinetics in muon catalyzed
fusion ($\mu$CF).
A method for determining the characteristics of MA processes in
$\mathrm{D}_2 +{}^{3}\mathrm{He}$ mixture is presented in the
following subsections.


\subsubsection{Muon stopping powers ratio}
\label{sec:muon-stopping-powers}

A muon entering the $\mathrm{D}_2 +{}^{3}\mathrm{He}$ mixture may be
captured by deuterium or helium into atomic orbits of excited $d\mu$
or $\mu\mathrm{He}$ atoms.
The corresponding relative probability has the following
form~\cite{kottm93,kottm87,koren90,bubak86,hartm90a}
\begin{equation}
  \label{eq10}
  W_\mathrm{D} = \frac{1}{1 + A\cdot c_{\mathrm{He}} }\, , \qquad
  W_{\mathrm{He}} = \frac{A\cdot c_{\mathrm{He}} }{1 + A\cdot
  c_{\mathrm{He}} }\, ,
\end{equation}
where $c_{\mathrm{He}} $ is the relative atomic helium concentration
in the $\mathrm{D}_2 +{}^{3}\mathrm{He}$ mixture and $A$ is the muon
stopping power ratio,
\begin{equation}
  \label{eq11}
  A = \frac{(dE/dx)_{\mathrm{He}} }{(dE/dx)_\mathrm{D}} \, ,
\end{equation}
with $(dE/dx)_{\mathrm{He}} $ and $(dE/dx)_\mathrm{D}$ the ionization
energy loss of muon per one atom of helium and deuterium.

The muon stopping power ratio $A$ may be experimentally determined
from the yield of electrons produced in muon decay processes in pure
helium and in $\mathrm{D}_2 +{}^{3}\mathrm{He}$ mixture.
However, the same constant momentum of the muonic beam must be kept
during expositions with both targets.
The experiment relies upon the determination of a density of a pure
helium target, $\tilde{\varphi}_{\mathrm{He}}$, (by variation of the
target density) such that the number of muon stops in the target (and
consequently the number of electrons arising from muons decaying from
the $d\mu$ and $^{3}\mathrm{He}$ ground state atoms) is the same as
the one obtained for a given $\mathrm{D}_2 +{}^{3}\mathrm{He}$
mixture.
Under this condition the moderation thickness of the pure helium
target is the same as the one of the $\mathrm{D}_2 +{}^{3}\mathrm{He}$
mixture with density $\varphi_{mix}$ and helium concentration
$c_{\mathrm{He}}$, given in Table~\ref{tab:exp-cond}, run III\@.
Using Eq.~(\ref{eq10}) one can determine the corresponding equivalent
density of the pure helium target as
\begin{equation}
  \label{eq13}
  A \cdot \tilde{\varphi}_{\mathrm{He}} = (c_\mathrm{D} + A\cdot
  c_{\mathrm{He}} )\varphi_{mix}\, ,
\end{equation}
where $\varphi_{mix}$ is the atomic density of the $\mathrm{D}_2
+{}^{3}\mathrm{He}$ mixture normalized to LHD and $c_\mathrm{D}$ and
$c_{\mathrm{He}}$ are the relative deuterium and helium concentrations
$(c_\mathrm{D} + c_{\mathrm{He}} = 1)$.

Analogously, one can use a pure deuterium target instead of the helium
one to determine the equivalent deuterium target density
$\tilde{\varphi}_\mathrm{D}$.
Then, equating numbers of electrons detected in runs with pure
deuterium and in the $\mathrm{D}_2 +{}^{3}\mathrm{He}$ target one can
obtain the equivalent density, $\tilde{\varphi}_\mathrm{D}$, and hence
the coefficient $A$
\begin{equation}
  \label{eq14}
  A = \frac{\tilde{\varphi}_\mathrm{D}-\varphi_{mix}c_\mathrm{D}}
  {\varphi_{mix} c_{\mathrm{He}} }\, .
\end{equation}
It should be noted, however, that a determination of $A$ from
Eq.~(\ref{eq14}) with the same accuracy as from Eq.~(\ref{eq13})
requires a significantly greater helium concentration in the
$\mathrm{D}_2 + {}^{3}\mathrm{He}$ mixture.


\subsubsection{The $\mathrm{q}_{1s}^{\mathrm{He}}$ probability}
\label{sec:prob-mathrmq_1s}

Prompt Lyman series transitions in $\mu\mathrm{He}$ atoms are also
observed in a $\mathrm{D}_2 +{}^{3}\mathrm{He}$ mixture.
As mentioned previously, they originate from direct muon capture by
deexcitation of $(\mu\mathrm{He})^*$ atoms or by muon transfer from
excited muonic deuterium to helium.
However, the relative intensities of $K$ series transitions measured
in a $\mathrm{D}_2 +{}^{3}\mathrm{He}$ mixture differ from the ones in
pure helium because effective reaction rates of $\mu\mathrm{He}$
deexcitation processes depend on the target conditions.

$\mathrm{q}_{1s}^{\mathrm{He}}$ represents the $(d\mu)^*$ atom
probability to reach the ground state in a $\mathrm{D}_2
+{}^{3}\mathrm{He}$ mixture and is defined as
\begin{equation}
  \label{eq15}
  \mathrm{q}_{1s}^{\mathrm{He}} = \frac{n_{d \mu}^{1s}}{n_{d \mu}^*}\,
   ,
\end{equation}
where $n_{d \mu}^*$ is the number of $d\mu$ atoms created in the
excited state due to direct muon capture in deuterium atoms, and
$n_{d \mu}^{1s}$ is the number of the $d\mu$ atoms which reach the
ground state during the cascade.
The number of $d\mu$ atoms created in the excited state can be written
as
\begin{equation}
  \label{eq16}
  n_{d \mu}^* = N_\mathrm{stop}^{\mathrm{D/He}} \cdot W_\mathrm{D}
\end{equation}
where $N_\mathrm{stop}^{\mathrm{D/He}}$ represents the number of muon
stops in the $\mathrm{D}_2 +{}^{3}\mathrm{He}$ gas mixture.

Since our setup is not able to measure $n_{d \mu}^{1s}$, we used
another method to determine $\mathrm{q}_{1s}^{\mathrm{He}}$.
The number of $\mu\mathrm{He}$ atoms formed in excited states due to
muon transfer from $(d\mu)^{*}$ to helium, $(d\mu)^* + \mathrm{He} \to
(\mathrm{He}\mu)^* +d$, is $n_{{\mathrm{He}}\mu^*}^{\mathrm{transf}}$
and corresponds to
\begin{equation}
  \label{eq17}
  n_{{\mathrm{He}}\mu^*}^{\mathrm{transf}} = n_{d \mu}^* - n_{d
  \mu}^{1s}\, .
\end{equation}
The total number of $\mu\mathrm{He}$ atoms created in the excited
states and emitting prompt $Kx$~lines is given by the yield
\begin{equation}
  \label{eq18}
  Y_{tot}^{\mathrm{D/He}} = \sum_{x= \alpha, \beta, \gamma}
  \frac{N^{\mathrm{D/He}}_x}{(1-\eta_x)\varepsilon_x} \, .
\end{equation}
On the other hand, $n_{{\mathrm{He}}\mu^*}^{\mathrm{dir}}$ is the
number of $\mu\mathrm{He}$ atoms formed in the excited states in
a $\mathrm{D}_2 + {}^{3}\mathrm{He}$ mixture due to direct muon capture
by helium atoms
\begin{equation}
  \label{eq19}
  n_{{\mathrm{He}}\mu^*}^{\mathrm{dir}} = Y_{tot}^{\mathrm{D/He}} -
  n_{{\mathrm{He}}\mu^*}^{\mathrm{transf}} =
  N_\mathrm{stop}^{\mathrm{D/He}} \cdot W_{\mathrm{He}}
\end{equation}
Isolating $n_{d \mu}^{1s}$ in Eq.~(\ref{eq17}), using
Eqs.~(\ref{eq16}) and~(\ref{eq19}), we obtain the
$\mathrm{q}_{1s}^{\mathrm{He}}$ probability as
\begin{equation}
  \label{eq20}
  \mathrm{q}_{1s}^{\mathrm{He}}= (1 + A\cdot c_{\mathrm{He}})\left [ 1
  -
  \frac{Y_{tot}^{\mathrm{D/He}}}{N_\mathrm{stop}^{\mathrm{D/He}}}\right
  ] \, .
\end{equation}
In the case of detecting events by the germanium detector in
coincidence with muon decay electrons, the total yield
$Y_{tot}^{\mathrm{D/He}}$ in Eq.~(\ref{eq20}) has to be replaced by
\begin{equation}
  \label{eq23}
  Y_{tot,\mathrm{e}}^{\mathrm{D/He}} = \frac{1}{\varepsilon_\mathrm{e}
  f_t} \sum \limits_{x = \alpha, \beta, \gamma}
  \frac{N^{\mathrm{D/He}}_{x-\mathrm{e}}}{(1-\eta_{x})\varepsilon_{x}}\,
  .
\end{equation}


\subsubsection{Radiative molecular peak}
\label{su:k}

The delayed muonic x~rays are generated by two different mechanisms
initiated by $d\mu$ atoms in their ground state.
The first mechanism described in this section is simply molecular muon
transfer, specifically Eq.~(\ref{eq1}a) accompanied by a 6.85~keV
$\gamma$--rays.
Experimental molecular muon transfer from muonic deuterium to helium
$\lambda_{d {}^3\mathrm{He}}$ is presented in detail in many papers,
in particular in Refs.~\cite{czapl96c,bystr03} together with the
corresponding reaction rates.
Below radiative decay rate of the $d\mu {}^{3}\mathrm{He}$ complex
Eq.~(\ref{eq1}a) can be measured as follows.

The time distribution of the $\gamma$~rays (relative to the muon stop
time) falls (in a pure target) experimentally off with the
disappearance rate of the muonic deuterium ground state, $\lambda_{d
\mu}$,
\begin{equation}
  \label{eq28}
  \frac{dN_{6.85}}{dt} = A_{d\mu} \cdot e^{-\lambda_{d \mu}t}\,
  ,
\end{equation}
with $A_{d\mu}$ the amplitude and 
\begin{eqnarray}
  \label{eq29}
  \lambda_{d \mu} & = & \lambda_0 + \lambda_{d^3\mathrm{He}}\varphi
  c_{\mathrm{He}} \nonumber \\
  & + & \tilde{\lambda}_{dd\mu}\varphi c_\mathrm{D} \left[1 -
  W_\mathrm{D} \mathrm{q}_{1s}^{\mathrm{He}}(1 - \beta \omega_d )
  \right ] \, .
\end{eqnarray}
$\lambda_{d^3\mathrm{He}}$ is the molecular formation rate for the
$d\mu{}^3\mathrm{He}$ molecule, $\lambda_0 = 0.455 \times 10^6 \,
\mbox{s}^{-1}$ is the free muon decay rate.
$\tilde{\lambda}_{dd\mu}$ is the effective $dd\mu$ molecule formation
rate, $\beta$ the relative probability of nuclear fusion in $dd\mu$
with neutron production in the final channel and $\omega_d $ is the
muon sticking probability to helium produced in nuclear $d-d$ fusion
(see~\cite{bystr03}).

The probability of the radiative decay of the $d\mu{}^{3}\mathrm{He}$
system (corresponding to the $2p\sigma \to 1s\sigma$ transition) is
defined by
\begin{equation}
  \label{eq30}
  \kappa_{d\mu\mathrm{He}} = \frac{\lambda_{\gamma}}{\lambda_{p} +
  \lambda_{\gamma} + \lambda_e}\, .
\end{equation}
where $\lambda_\gamma$, $\lambda_p$, and $\lambda_e$ are the reaction
rates for the $d\mu {}^{3}\mathrm{He}$ molecular decay according to
channels (a), (b) and (c) of Eq.~(\ref{eq1}), respectively, also shown
in Fig.~\ref{fig:2}.
The formation of the $d\mu {}^{3}\mathrm{He}$ molecule practically
coincides with the subsequent $\gamma$--ray emission because of the
very short average life--time of $d\mu {}^{3}\mathrm{He}$ molecule
($\sim 10^{11} \,
\mbox{s}^{-1}$~\cite{bystr95d,augsb03,tresc98c,gartn00,kinox93}).

In the present experiment only the radiative decay channel is
detected.
The corresponding $\kappa_{d\mu\mathrm{He}}$ probability is determined
by the ratio
\begin{equation}
  \label{eq31}
  \kappa_{d\mu\mathrm{He}} = \frac{N_\gamma^{d \mu^3{\mathrm{He}} }}
  {N_{tot}^{d \mu^3{\mathrm{He}} }}\, ,
\end{equation}
where $N_{tot}^{d \mu^3{\mathrm{He}}}$ and $N_\gamma^{d
\mu^3{\mathrm{He}}}$ are the total number of $d\mu {}^{3}\mathrm{He}$
molecules formed in the mixture and the number of molecules
subsequently decaying via the radiative channel, Eq.~(\ref{eq1}a),
respectively).
The latter quantity may be expressed as 
\begin{equation}
  \label{eq32}
  N_\gamma^{d \mu^3{\mathrm{He}} } =
  \frac{N_{6.85}}{\varepsilon_{6.85} F_t (1-\eta_{6.85})}\, ,
\end{equation}
where $N_{6.85}$ is the number of 6.85~keV $\gamma$-rays during time
$\Delta t_\gamma$ elapsed after a muon stop and $\varepsilon_{6.85}$
is the corresponding detection efficiency.
The factor $F_t$
\begin{equation}
  \label{eq33}
  F_t = e^{-\lambda_{d \mu} t}(1 - e^{-\lambda_{d \mu} \Delta
  t_\gamma})
\end{equation}
is the $\gamma$--ray detection time factor and $\eta_{6.85}$ is the
6.85~keV $\gamma$--ray attenuation coefficient.
For the $\gamma$~rays detected with the del-$e$ criterion, a
corresponding $N_\gamma^{d \mu^3{\mathrm{He}}}$ value is obtained
using Eq.~(\ref{eq32}) divided by the $\varepsilon_\mathrm{e} f_t$
coefficients.

A comparison of the $N_\gamma^{d \mu^3{\mathrm{He}}}$ value measured
with and without del-$e$ criterion provides also a test for the
validity of our coefficients $\varepsilon_\mathrm{e}$, $f_t$, and
$N_{6.85}$.
The detection efficiency $\varepsilon_{6.85}$ was determined by MC
simulations including feasible space distributions of muon stops in
the target volume and experimental detection efficiencies of
$Kx$~lines for the pure ${}^{3}\mathrm{He}$ runs.

The total number of the $d\mu {}^{3}\mathrm{He}$ molecules formed in
$\mathrm{D}_{2}+\mathrm{He}$ mixture is determined by analyzing the
6.85~keV $\gamma$--ray time distribution.
It is expressed as
\begin{equation}
  \label{eq34}
  N_{tot}^{d \mu^3{\mathrm{He}} } = \frac{\lambda_{d {}^3
  {\mathrm{He}} } \varphi \mathrm{c}_{\mathrm{He}} }{\lambda_{d \mu}}
  n_{d\mu}^{1s}\, ,
\end{equation}
where $n_{d\mu}^{1s}$ is the number of $d\mu$ atoms formed via direct
muon capture and reached the ground state after deexcitation.
By measuring the exponential time distribution (\ref{eq28}) and using
the known quantities $\lambda_0,\ \tilde{\lambda}_{dd\mu},\
W_\mathrm{D},\ \omega_d ,\ \mathrm{q}_{1s}^{\mathrm{He}},\ \beta$
\cite{balin84b,balin84,vorob88} one can determine the molecular
formation rate $\lambda_{d {}^3{\mathrm{He}}}$ from Eq.~(\ref{eq29}).
The determination of $N^{tot}_{d \mu^3{\mathrm{He}}}$ from
Eq.~(\ref{eq34}) requires in addition the knowledge of
$n_{d\mu}^{1s}$, isolated in Eqs.~(\ref{eq15}) and~(\ref{eq16}).
By substituting $N_\gamma^{d \mu^3{\mathrm{He}}}$ and $N_{tot}^{d
\mu^3{\mathrm{He}}}$ into Eq.~(\ref{eq31}) one finally obtains the
$\kappa_{d\mu\mathrm{He}}$ probability.


\subsubsection{Delayed $K$ series transitions from muonic helium}
\label{sec:delayed-k-series}

As already said, the delayed muonic x~rays are generated by two
different mechanisms initiated by the ground state $d\mu$ atoms.
The second one discussed here is proceeded by $dd\mu$ formation (in
collision of $(d\mu)_{1s}$ with $\mathrm{D}_2$ molecule) and
subsequently by nuclear $d-d$ fusion.
Muons freed after fusion form excited muonic helium atoms due to
direct muon capture by helium or due to muon capture by deuterium and
subsequent muon transfer to helium.
Then the delayed x~rays of muonic helium $K$ series transitions are
observed.

The time distribution is also determined by $\lambda_{d \mu}$.
Besides, the relative intensities $I_{x,del}$ (or $I_{x-e,del}$) of
the delayed $K$ series transitions are assumed to be the same as those
of the prompt radiation of $Kx$~lines.
It is worthwhile to note that the measurement of the corresponding
absolute intensities enabled us to determine the third component of
$\lambda_{d \mu}$ in Eq.~(\ref{eq29}) and, consequently, to extract
the effective formation rate of the $dd\mu$ molecule in the
$\mathrm{D}_2 +{}^{3}\mathrm{He}$ mixture using the coefficients
$W_\mathrm{D},\ \mathrm{q}_{1s}^{\mathrm{He}}$ (also obtained in this
paper) and average values for $\beta$, and $\omega_d$ (taken from
Refs.~\cite{balin84b,balin84,vorob88}).


\section{Analysis}
\label{sec:analysis}


\subsection{Relative intensities of $K$ series transitions }
\label{sec:meas-relat-intens}

\begin{figure}[ht]
\includegraphics[width=5.cm,angle=90]{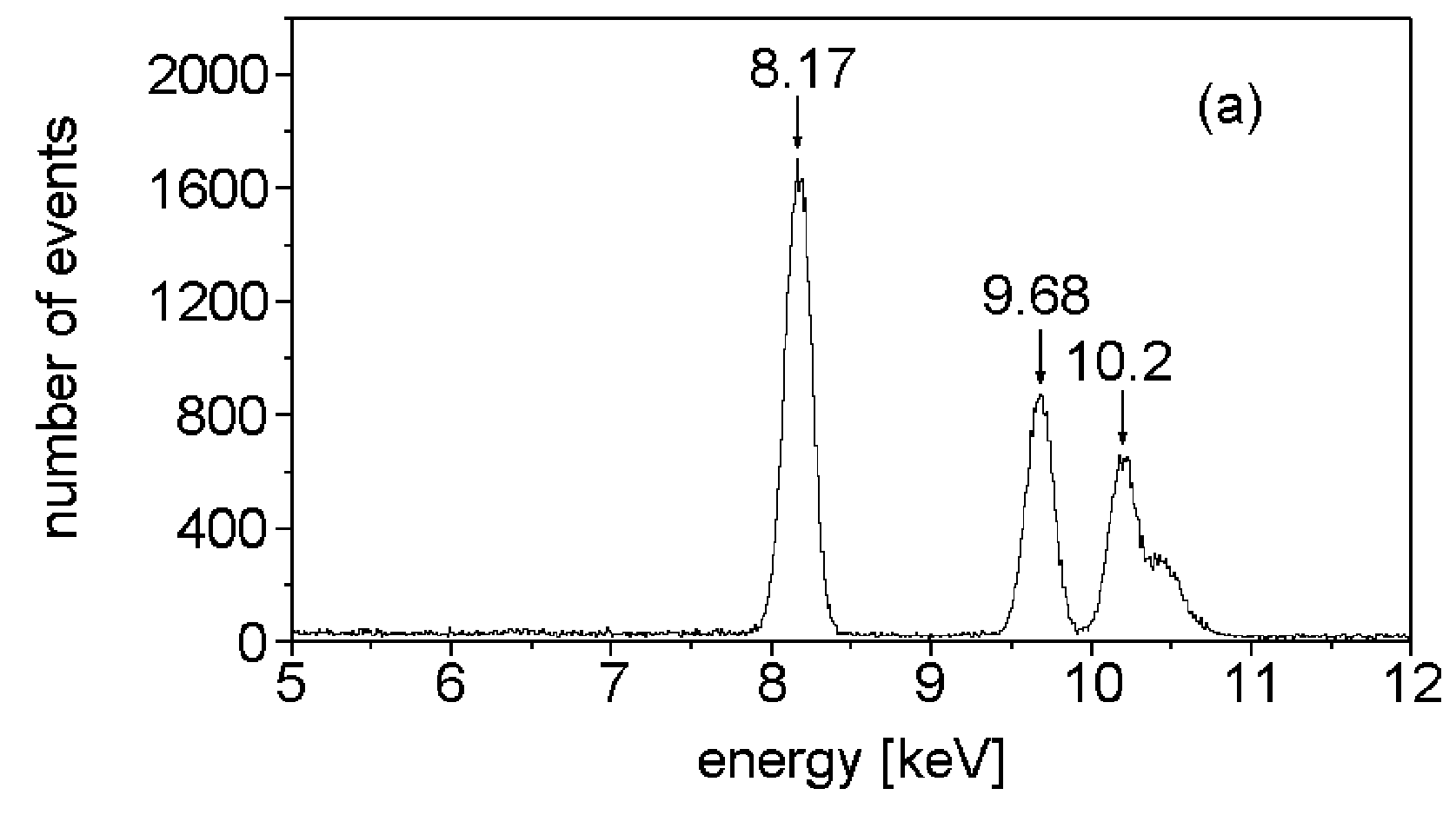}
\includegraphics[width=5.cm,angle=90]{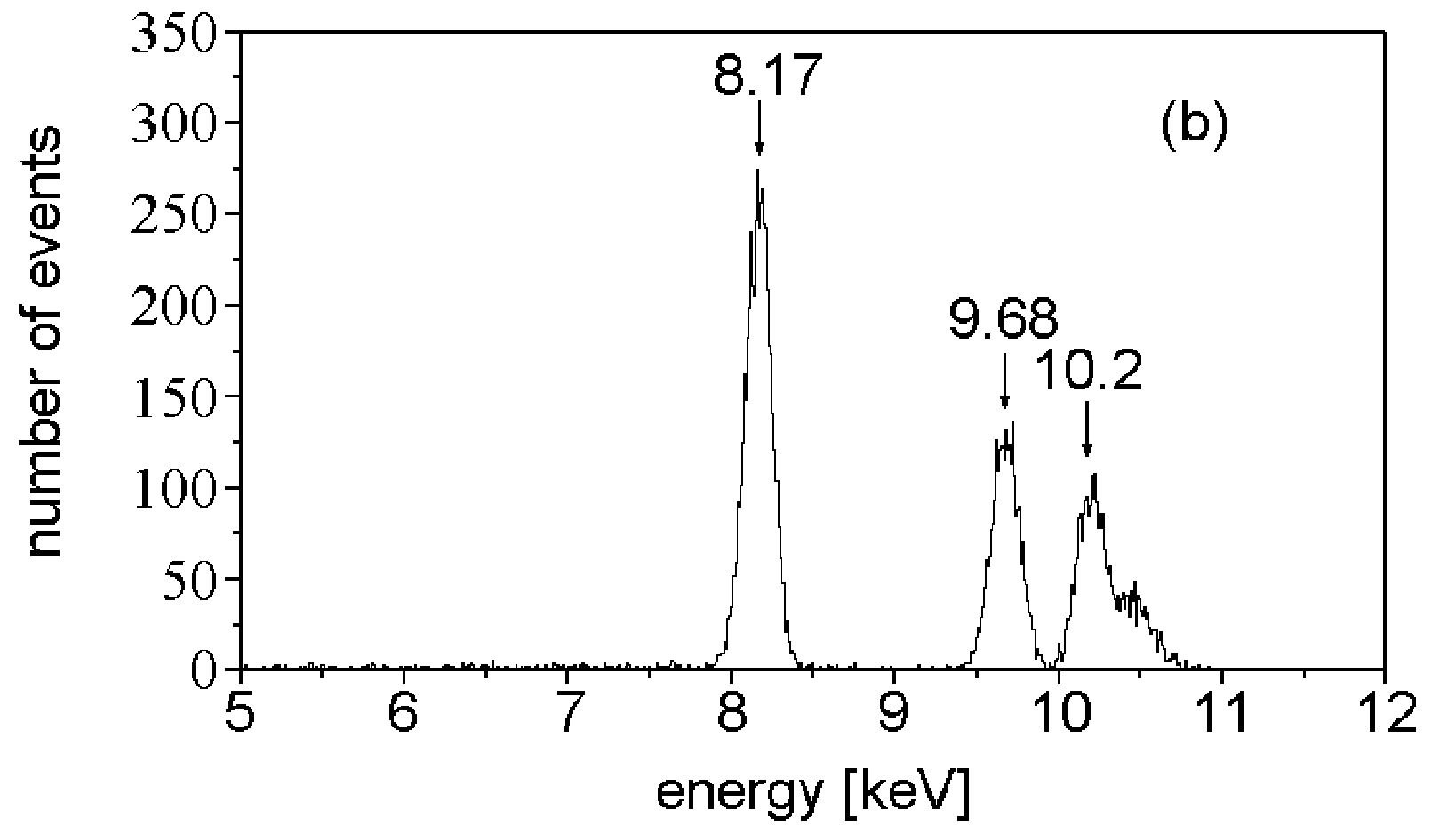}
                \caption{Prompt events energy distribution in 
                        run I without (a) and with coincidences with
                        muon decay electrons (b).}
\label{fig:ene-prompt}
\end{figure}

\begin{figure}[ht]
\includegraphics[width=5.cm,angle=90]{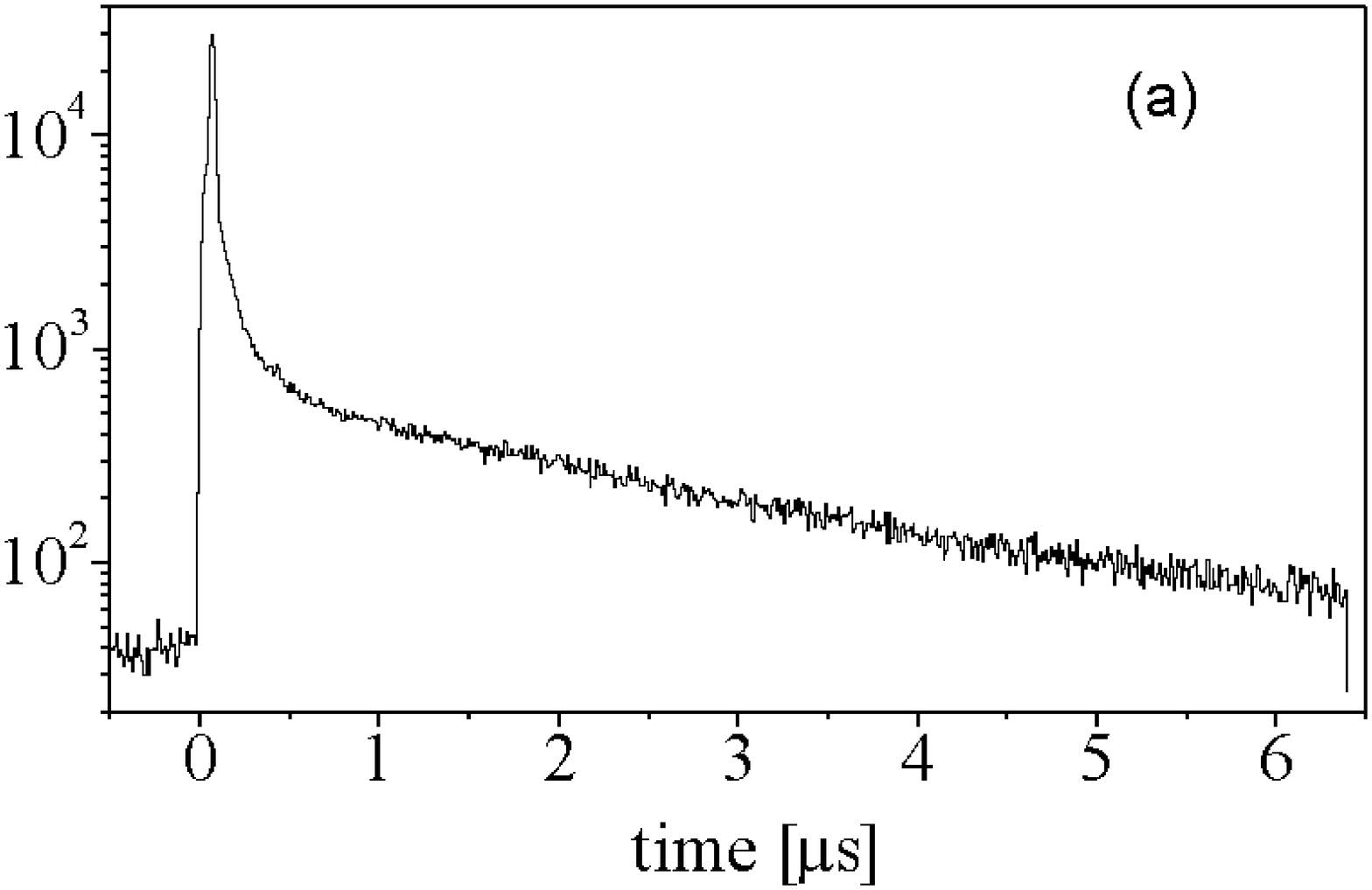}
\includegraphics[width=5.cm,angle=90]{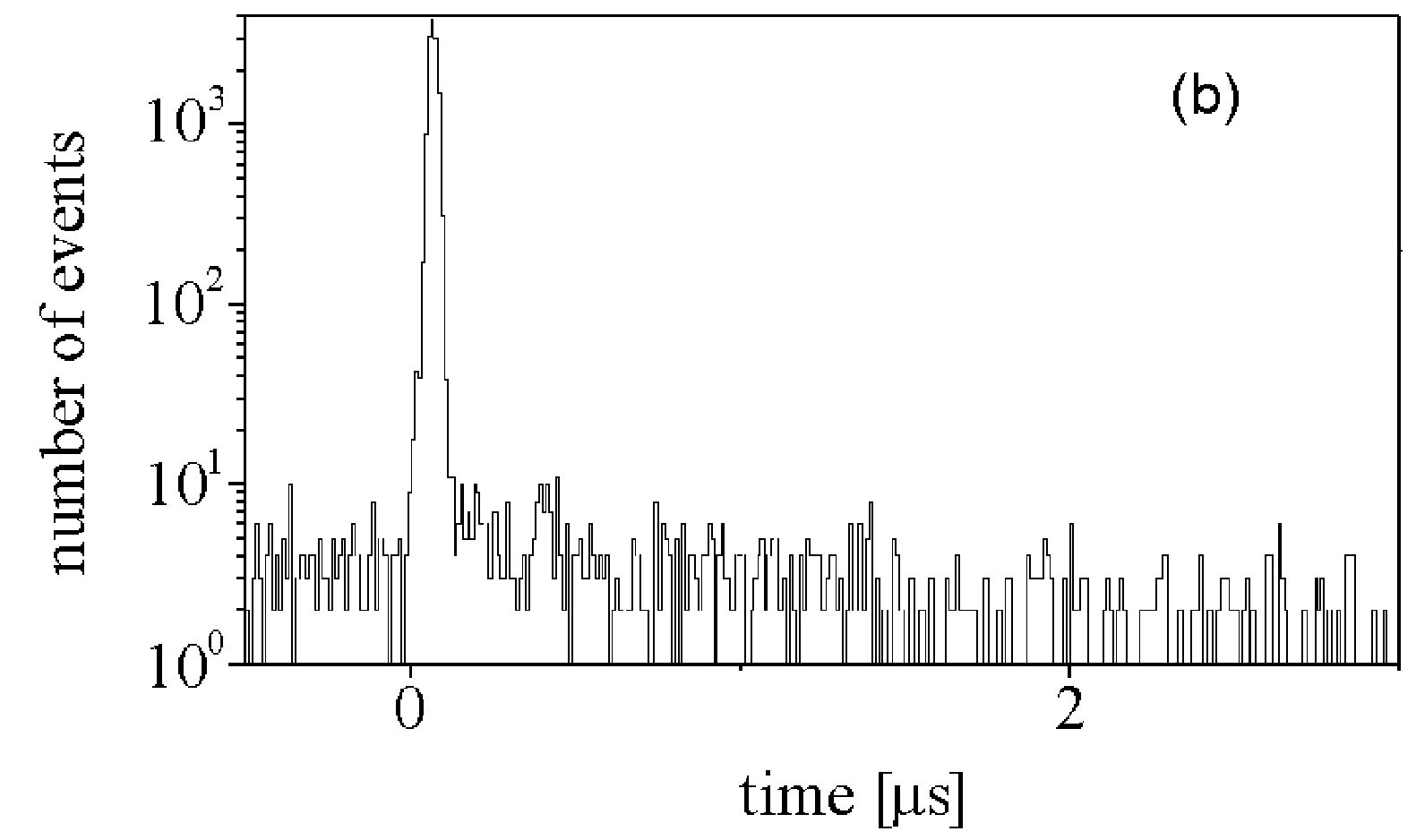}
                \caption{Time distribution in run I without (a) and
                        with coincidences with muon decay electrons
                        (b).}
\label{fig:time-ges}
\end{figure}

\begin{table*}[ht]
\begin{ruledtabular}
       \caption{Prompt x--ray yields of $\mu{}^{3,4}\mathrm{He}$
               $K$~series transitions measured in different runs with
               pure ${}^{3}\mathrm{He}$ and ${}^{4}\mathrm{He}$. }
\label{tab:yield}
\begin{tabular}{cccccccccc}
  & \multicolumn{2}{c}{$K\alpha$} & \multicolumn{2}{c}{$K\beta$} &
\multicolumn{2}{c}{$K\gamma$} &  \multicolumn{3}{c}{Yield}\\
Range [keV] & \multicolumn{2}{c}{$[7.83-8.53]$} &
\multicolumn{2}{c}{$[9.43-9.96]$} & \multicolumn{2}{c}{$[9.98-10.6]$}
& $[10^{8}]$ & $[10^{8}]$ & $[10^{8}]$ \\
Runs & $N^{\mathrm{He}}_\alpha$ &
$N^{\mathrm{He}}_{\alpha-\mathrm{e}}$ & $N^{\mathrm{He}}_\beta$ &
$N^{\mathrm{He}}_{\beta-\mathrm{e}}$ & $N^{\mathrm{He}}_\gamma $ &
$N^{\mathrm{He}}_{\gamma-\mathrm{e}}$ & $Y^{\mathrm{He}}_{\alpha}$ &
$Y^{\mathrm{He}}_{\beta}$ & $Y^{\mathrm{He}}_{\gamma}$ \\ \hline
I (${}^{3}\mathrm{He}$) & 34 319(190) & 4785(70)& 17 835(139)&
2551(52)& 20 045(150)& 2834(54)& 7.536(90) & 3.795(53) & 4.231(62) \\
IIa (${}^{4}\mathrm{He}$) & 7295(87) & 985(32) & 4919(72) & 688(26) &
2616(55) & 408(20) & 0.897(14) & 0.585(10) & 0.309(8) \\
IIb (${}^{4}\mathrm{He}$) & 11 587(111) & 1593(40)& 7547(91) &
1009(32)& 4627(76) & 613(25) & 1.766(25) & 1.126(18) & 0.677(13) \\
IIc (${}^{4}\mathrm{He}$) & 1303(38) & 174(14) & 709(29) & 91(10) &
846(33) & 123(12) & 0.287(9) & 0.151(6) & 0.178(7) \\
\end{tabular}
\end{ruledtabular}
\end{table*}

\begin{table*}[ht]
\begin{ruledtabular}
       \caption{Relative intensities of prompt x~rays of
                $\mu{}^{3,4}\mathrm{He}$ $K$ series transitions
                measured in runs with pure helium. For each run,
                results from both the full statistics and the del-$e$
                condition are given. }
\label{tab:intens}
\begin{tabular}{ccccccc}
 & $I^{\mathrm{He}}_\alpha$ & $I^{\mathrm{He}}_{\alpha - \mathrm{e}}$
 & $I^{\mathrm{He}}_\beta$ & $I^{\mathrm{He}}_{\beta -\mathrm{e}}$ &
 $I^{\mathrm{He}}_\gamma$ & $I^{\mathrm{He}}_{\gamma -\mathrm{e}}$ \\
& [\%] & [\%] & [\%] & [\%] & [\%] & [\%] \\ \hline
I (${}^{3}\mathrm{He}$) & 48.4(8) & 47.8(9) & 24.4(4) & 24.8(6) &
27.2(4) & 27.4(6) \\
IIa (${}^{4}\mathrm{He}$) & 50.0(9) & 47.3(19) & 32.7(7) & 33.1(14) &
17.3(5) & 19.6(13) \\
IIb (${}^{4}\mathrm{He}$) & 49.5(9) & 49.5(15) & 31.5(6) & 31.4(11) &
19.0(5) & 19.1(8) \\
IIc (${}^{4}\mathrm{He}$) & 46.6(18)& 44.8(44)& 24.5(11)& 23.5(29)&
28.9(13)& 31.7(35)\\
Augsburger \textit {et al.}~\cite{augsb03} (${}^{4}\mathrm{He}$) &
46.9(45)& --- & 27.9(28) & --- & 25.2(19)& --- \\
Tresch \textit {et al.}~\cite{tresc98c}\footnotemark[1] & 47.0(2) &
--- & 20.3(10) & --- & 32.7(16)& --- \\
\end{tabular}
            \footnotetext[1]{for ${}^{3}\mathrm{He}$ ($\varphi=0.026$)
                         and for ${}^{4}\mathrm{He}$
                         ($\varphi=0.0395$)}
\end{ruledtabular}
\end{table*}

To obtain the relative intensities of muonic x--ray $K$ series
transitions of $\mu{}^3\mathrm{He}$ and $\mu{}^4\mathrm{He}$ atoms in
helium targets we analyzed the corresponding energy and time
distributions detected by the germanium detector in runs I and II\@.
Figures~\ref{fig:ene-prompt} and~\ref{fig:time-ges} present the energy
and time distributions obtained in runs I with and without muon
decay electrons coincidences.
As seen, the del-$e$ criterion significantly suppresses the
background level and improved the signal--to--background ratio.
As already mentioned before, events detected within a time interval
$t_\gamma = [(-0.03) - (+0.03)] \, \mu$s relative to muon stops were
classified as prompt ones.
The prompt $Kx$~lines events $N^{\mathrm{He}}_x,\
N^{\mathrm{He}}_{x-\mathrm{e}}$ were determined by fitting the
experimental amplitude distributions by a Gaussian distribution
\begin{equation}
  \label{eq37}
  \frac{dN^{\mathrm{He}}_x}{d \mathrm{E}_x} = A_x \cdot \exp \left [
  {-\frac{(\mathrm{E}_x-\overline{\mathrm{E}}_x)^2} {2\sigma^2_x}}
  \right ] + S \cdot \mathrm{E}_x + O\, ,
\end{equation}
where $\overline{\mathrm{E}}_x$ is the mean value of the corresponding
$Kx$~line energy, $\sigma_x$ the standard deviation for the $Kx$~line
and $A_x$ the normalization constant.
The germanium detector background is taken into account by a straight
line, with $S$ and $O$ being the constants.
Results obtained in measurements I and II are presented in
Tables~\ref{tab:yield} and~\ref{tab:intens}.
Statistical errors are quoted in parentheses throughout the whole
text.

The analysis performed for both mixtures is similar.
The prompt intensities are measured within the same time interval as
for the pure helium runs, both with and without the delayed electron
coincidence condition.
The results, given in Table~\ref{tab:intens-mix}, depend on the
pressure of the $\mathrm{D}_2 + {}^3\mathrm{He}$ mixture.
For comparison, results of Augsburger \textit{et al.}~\cite{augsb03}
taken at a similar pressure as in run III, are also shown in the table.
The differences in relative intensity between pure helium and the
deuterium--helium mixtures are essentially due to excited state
transfer.
Additionally, such an analysis allows us to determine the $Kx$
transition energy differences between the two helium isotopes.
The $\Delta E({}^4 \mathrm{He}-{}^3 \mathrm{He})$ energy differences
are given in Table~\ref{tab:isotop}.
A theoretical prediction exists for the $K\alpha$
transition~\cite{rinke76} which is slightly lower than our measured
value.

\begin{table}[hb]
\begin{ruledtabular}
       \caption{$Kx$ transition energy differences between the two
       helium isotopes. The last column gives a theoretical prediction
       for the $K\alpha$ transition.}
\label{tab:isotop}
\begin{tabular}{cccc}
Transitions & \multicolumn{3}{c}{$\Delta E({}^4 \mathrm{He}-{}^3
\mathrm{He})$ [eV]} \\
& Our work & Tresch \textit{et al.}~\cite{tresc98c} &
Rinker~\cite{rinke76} \\ \hline
$K\alpha$ & $77.8 \pm 0.9$ & $75.0 \pm 1.0$ & 74.2\\
$K\beta$ & $92.9 \pm 1.1$ & --- \\
$K\gamma$ &$103.4 \pm 3.4$& ---  \\
\end{tabular}
\end{ruledtabular}
\end{table}

\begin{table*}[ht]
\begin{ruledtabular}
       \caption{Relative intensities, in percent, of prompt x~rays of
                $\mu{}^3\mathrm{He}$ $K$ series transitions measured
                in runs III and IV\@.  ``Full'' stands for full
                statistics, whereas del-$e$ represents the delayed
                electron criterion. The last column presents the
                results of Augsburger \textit {et
                al.}~\cite{augsb03}.}
\label{tab:intens-mix}
\begin{tabular}{ccccccc}
Runs & \multicolumn{2}{c}{III} & \multicolumn{2}{c}{IV} & Augsburger
\textit {et al.}~\cite{augsb03} \\
Transitions & full & del-$e$ & full & del-$e$ \\\hline
$I^{\mathrm{D/He}}_\alpha$ & 66.4(7) & 65.7(15) & 72.0(6) & 72.9(16) &
68.6(51) \\
$I^{\mathrm{D/He}}_\beta$ & 26.6(5) & 26.5(8) & 24.5(3) & 24.1(8) &
24.5(19) \\
$I^{\mathrm{D/He}}_\gamma$ & 7.0(4) & 7.8(4) & 3.5(1) & 3.0(3) &
6.9(6) \\
\end{tabular}
\end{ruledtabular}
\end{table*}


\subsection{ Stopping power ratio measurement}
\label{su:A}

The stopping power ratio $A$ was determined by analyzing the time
spectra of muon decay electrons measured by the scintillator pairs in
the runs with pure helium (run I) and the $\mathrm{D}_2
+{}^{3}\mathrm{He}$ mixture (run III).
Measurements with pure helium were performed to find a corresponding
equivalent density $\tilde{\varphi}_{\mathrm{He}}$ (see
Eq.~(\ref{eq13}) resulting in the same number of muon stops in the
target volume as the one in the $\mathrm{D}_2 +{}^{3}\mathrm{He}$
mixture (run III).
Variation of the helium target density enabled us to find a density
dependence of muon stops and hence $\tilde{\varphi}_{\mathrm{He}}$.
The $A$ stopping power ratio was the determined according to
Eq.~(\ref{eq13}).
Note that the linear density dependence of muon stops in gases is a
good approximation, if the gaseous mixture thickness expressed in
terms of energy is significantly greater than the energy dispersion of
the muon beam.
This condition is fulfilled in our experiments.

Due to muons stopping also in aluminum and gold (target walls) besides
in the gas, the electron time spectra are a sum of exponential
functions:
\begin{eqnarray}
  \label{eq38}
      \frac{dN_e }{dt} & = & A^e_{\mathrm {Al}} \cdot e^{ - \lambda
      _{\mathrm {Al}} \cdot t} + A^e_{\mathrm {Au}} \cdot e^{ -
      \lambda _{\mathrm {Au}} \cdot t} \nonumber \\
      & + & A^e_{\mathrm {He}} \cdot e^{ - \lambda _{\mathrm {He}}
      \cdot t} + B^e,
\end{eqnarray}
where $A^e_{\mathrm {Al}}$, $A^e_{\mathrm {Au}}$, and $A^e_{\mathrm
{He}}$ are the corresponding normalization amplitudes and
\begin{eqnarray}
      \label{eq38a}
      \lambda _{\mathrm {Al}} & = & Q_{\mathrm {Al}} \cdot \lambda_0 +
      \lambda_\mathrm{cap}^{\mathrm {Al} } \nonumber \\
      \lambda _{\mathrm {Au}} & = & Q_{\mathrm {Au}} \cdot \lambda_0 +
      \lambda_\mathrm{cap}^{\mathrm {Au} } \\
      \lambda _{\mathrm {He}} & = & \lambda_0 +
      \lambda_\mathrm{cap}^{\mathrm {He} }, \nonumber
\end{eqnarray}
are the muon disappearance rates in the different elements (the rates
are the inverse of the muon lifetimes in the target wall materials).
The nuclear capture rates in aluminum and gold,
$\lambda_\mathrm{cap}^{\mathrm {Al}}= 0.7054 (13) \times
10^6$~s$^{-1}$ and $\lambda_\mathrm{cap}^{\mathrm {Au}}= 13.07 (28)
\times 10^6$~s$^{-1}$, are known~\cite{suzuk87}.
$Q_{\mathrm{Al}}$ and $Q_{\mathrm{Au}}$ are the Huff factors, which
take into account that muons are bound in the $1s$ state of the
respective nuclei when they decay.
This factor is negligible for helium but necessary for aluminum
$Q_{\mathrm {Al}}=0.993$ and important for gold
$Q_{\mathrm{Au}}=0.850$~\cite{suzuk87}.
The constant $B^e$ characterizes the random coincidence background.

By measuring the helium amplitude, $A^e_{\mathrm {He}}$, 
\begin{equation}
      \label{eq39}
      A^e_{\mathrm {He}} = N_{\mathrm{stop}}^{\mathrm{He}} \cdot
      \varepsilon _e \cdot \lambda _0
\end{equation}
and knowing the electron detection efficiencies averaged over the
energy distributions ($\varepsilon _e$), one obtains the number of
muons stopping in helium $N_{\mathrm{stop}}^{\mathrm{He}}$.

\begin{figure}[ht]
\includegraphics[width=5cm,angle=90]{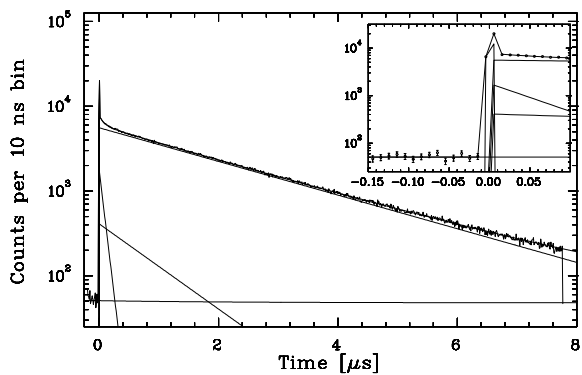}
                \caption{Time distribution of muon-decay electrons
                        measured in run I\@.  The inset shows details
                        at early times.}
\label{fig:telec}
\end{figure}

Figure~\ref{fig:telec} shows the time distribution of the muon decay
electrons measured in run I with the three exponential fit
(Eq.~(\ref{eq38})).
The ratio $R$ of muons stopped in the target, i.e., the number of
detected electrons $N_e$, divided by the number of muons entering the
target $N_\mu$
\begin{equation}
  \label{eq40}
  R(\varphi) = \frac{N_e}{N_\mu} = a\varphi + b
\end{equation}
depends on the helium density.
Table~\ref{tab:mudecay} shows the measured values of $R$ in percent
for each measurements. 
By fitting $R$ as a function of $\varphi$, one obtains
\begin{equation}
  \label{eq40a}
  a = -(0.60 \pm 0.11) \qquad b = (0.099 \pm 0.005)
\end{equation}
The value of $\tilde{\varphi}_{\mathrm{He}}$ was determined from the
condition
\begin{equation}
  \label{eq41}
  R(\tilde{\varphi}_{\mathrm{He}}) = R({\mathrm{D/He}})\, ,
\end{equation}
where $R(\tilde{\varphi}_{\mathrm{He}})$ and $R({\mathrm{D/He}})$ are
the ratios for pure ${}^{3}\mathrm{He}$ (at the density
$\tilde{\varphi}_{\mathrm{He}}$) and for the $\mathrm{D}_2
+{}^{3}\mathrm{He}$ mixture, respectively.
We found
\begin{equation}
  \label{eq41a}
  \tilde{\varphi}_{\mathrm{He}} = 0.0361 \left [
  ^{+0.0084}_{-0.0057}\right ] \, .
\end{equation}
The corresponding value of $A$, using Eq.~(\ref{eq13}), is then
\begin{equation}
  \label{eq42}
  A = 1.67 \left [ ^{+0.35}_{-0.33} \right ]\, .
\end{equation}

\begin{table}[hb]
\begin{ruledtabular}
      \caption{$R$ ratio measurements for runs I and run III\@.
              $N_\mu$ is the number of muons entering the target and
              $N_e$ the number of detected electrons. The beam
              momentum was $p_\mu = 34.0$~MeV/c.}
\label{tab:mudecay}
\begin{tabular}{ccccc}
Run & $N_\mu$ & $N_e$ & $R$ & $\varphi$ \\
    & $[10^6]$ & $[10^6]$ & $[10^{-2}]$ & [LHD] \\ \hline
Ia  & 1362.5 & 103.8(4)& 7.62(3) & 0.0363  \\
Ib  &  704.3 & 54.3(3) & 7.72(4) & 0.0359 \\
Ic  &  750.7 & 58.2(3) & 7.75(4) & 0.0355 \\
Id  &  413.6 & 32.5(2) & 7.85(6) & 0.0337 \\
III   & 8875.2 & 683.6(11) & 7.70(1) & 0.0585 \\
\end{tabular}
\end{ruledtabular}
\end{table}

The muon stopping power ratio $A$ of helium to deuterium atoms
coincides (within the experimental errors) with the results of
Refs.~\cite{kottm93,bystr93d} obtained under quite different
experimental conditions.
Our rather large relative errors, $\approx 20\%$, are a consequence of
small statistics and the relatively small range of variations of
${}^{3}\mathrm{He}$ and $\mathrm{D}_2 +{}^{3}\mathrm{He}$ target
densities.


\subsection{$\mathrm{q}_{1s}^{\mathrm{He}}$ probability}
\label{sec:meas-mathrmq_1s-para}

On of the main aim of runs III and IV was a measurement of the
$\mathrm{q}_{1s}^{\mathrm{He}}$ probability.
In order to determine this quantity it was necessary to know
(according to Eqs.~(\ref{eq15})-(\ref{eq20})) the muon stopping power
ratio $A$, the prompt $K$ series transition yields of
$\mu{}^3\mathrm{He}$ atoms in pure ${}^{3}\mathrm{He}$ and in
$\mathrm{D}_2 +{}^{3}\mathrm{He}$ mixtures, $N^{\mathrm{He}}_x$ and
$N^{\mathrm{D/He}}_x$, and the numbers of muon stops in pure
${}^{3}\mathrm{He}$ and in $\mathrm{D}_2 +{}^{3}\mathrm{He}$ mixtures,
$N_{\mathrm{stop}}$.
Significant background reduction was achieved by using the del-$e$
criterion.
The results are presented in Table~\ref{tab:expval}.
Note the excellent agreement between full statistics and del-$e$
analysis.

\begin{table}[ht]
\begin{ruledtabular}
      \caption{Experimental values of $\mathrm{q}_{1s}^{\mathrm{He}}$
              obtained from the $\mathrm{D}_2+ {}^{3}\mathrm{He}$
              experiments. ``Full'' stands for the full statistics,
              whereas del-$e$ represent the delayed electron
              criterion.}
\label{tab:expval}
\begin{tabular}{ccccc}
Runs & Statistics & $\sum\limits_{x = \alpha, \beta, \gamma}
N^{\mathrm{D/He}}_x$ & $Y_{tot}^{\mathrm{D/He}}$ &
$\mathrm{q}_{1s}^{\mathrm{He}}$ \\
    &            &              & $[10^8]$   &             \\ \hline
III & full       & 35 376(270)  &  7.70(15) & 0.882(18)  \\
    & del-$e$    & 4968(72)    &  7.60(29) & 0.885(21)  \\
IV  & full       & 37 402(205)  &  5.71(11) & 0.844(20)  \\
    & del-$e$    & 5161(75)    &  5.85(23) & 0.838(23)  \\
\end{tabular}
\end{ruledtabular}
\end{table}

\begin{figure}[ht]
\includegraphics[width=5.5cm,angle=90]{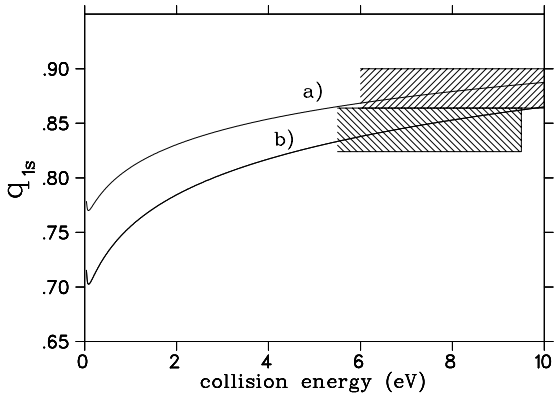}
                \caption{Energy dependence of theoretical
		  $\mathrm{q}_{1s}^{\mathrm{He}}$ in $\mathrm{D}_2 +
		  {}^3 \mathrm{He}$ mixture calculated for runs~III
		  (curve a) and~IV (curve b). Experimental values of
		  the $\mathrm{q}_{1s}^{\mathrm{He}}$ measured in the
		  present work ($\mathrm{q}_{1s}^{\mathrm{He}} =
		  (0.882\pm 0.018)$ and $\mathrm{q}_{1s}^{\mathrm{He}}
		  = (0.844\pm 0.020)$) are represented by hatched
		  boxes defined by their values and errors.}
\label{fig:calcq}
\end{figure}

Figure~\ref{fig:calcq} shows the energy dependence of the theoretical
$\mathrm{q}_{1s}^{\mathrm{He}}$ values versus $d\mu + {}^3\mathrm{He}$
collision energy calculated for runs~III and~IV in the framework of the
simple $(d \mu)^*$ cascade model~\cite{bystr99,bystr00,bystr99e} and
compares then with experiment.
The model assumes that the kinetic energy of $(d \mu)^*$ atoms remains
unchanged during deexcitation.
The $\mathrm{q}_{1s}^{\mathrm{He}}$ value is determined from
deexcitation and muon transfer to helium.
A complicated interplay between these two processes is described by a
system of linear first order differential equations for level
populations, $N_{nl}(t)$, with $n \leq 12$.
The $\mathrm{q}_{1s}^{\mathrm{He}}$ is defined as
\begin{equation}
  \label{eq42a}
  \mathrm{q}_{1s} = N_{1s}(t\rightarrow \infty) \,.
\end{equation}
The deexcitation scheme is taken from Ref.~\cite{bystr99} and the
corresponding reaction rates are collected in
Refs.~\cite{bystr99,bystr00}.

As seen from Fig.~\ref{fig:calcq}, the experimental values of
$\mathrm{q}_{1s}^{\mathrm{He}}$ coincide with the theoretical ones for
an average $d\mu$--He collision energy around 8~eV\@.
Note the pronounce difference between the experimental values of
$\mathrm{q}_{1s}^{\mathrm{He}}$ and the theoretical ones corresponding
to fully thermalized $d\mu$ atoms.
However, more refined theoretical calculations of
$\mathrm{q}_{1s}^{\mathrm{He}}$ based on Monte Carlo simulations of
acceleration of $d\mu$ atoms due to deexcitation processes and muon
transfer to helium as well as thermalisation due to elastic collisions
are required to arrive at definite conclusions.
It should also be noted that experimental results presented in this
paper agree with earlier ones (see Ref.~\cite{bystr90}).
On the other hand, analogous comparison with results presented in
Refs.~\cite{kottm93,augsb03,tresc98c,kottm87} is not possible due to
significantly different helium concentrations and densities.


\subsection{Radiative branching ratio $\kappa_{d\mu\mathrm{He}}$}
\label{sec:radi-branch-ratio}

\begin{figure}[ht]
\includegraphics[width=5.cm,angle=90]{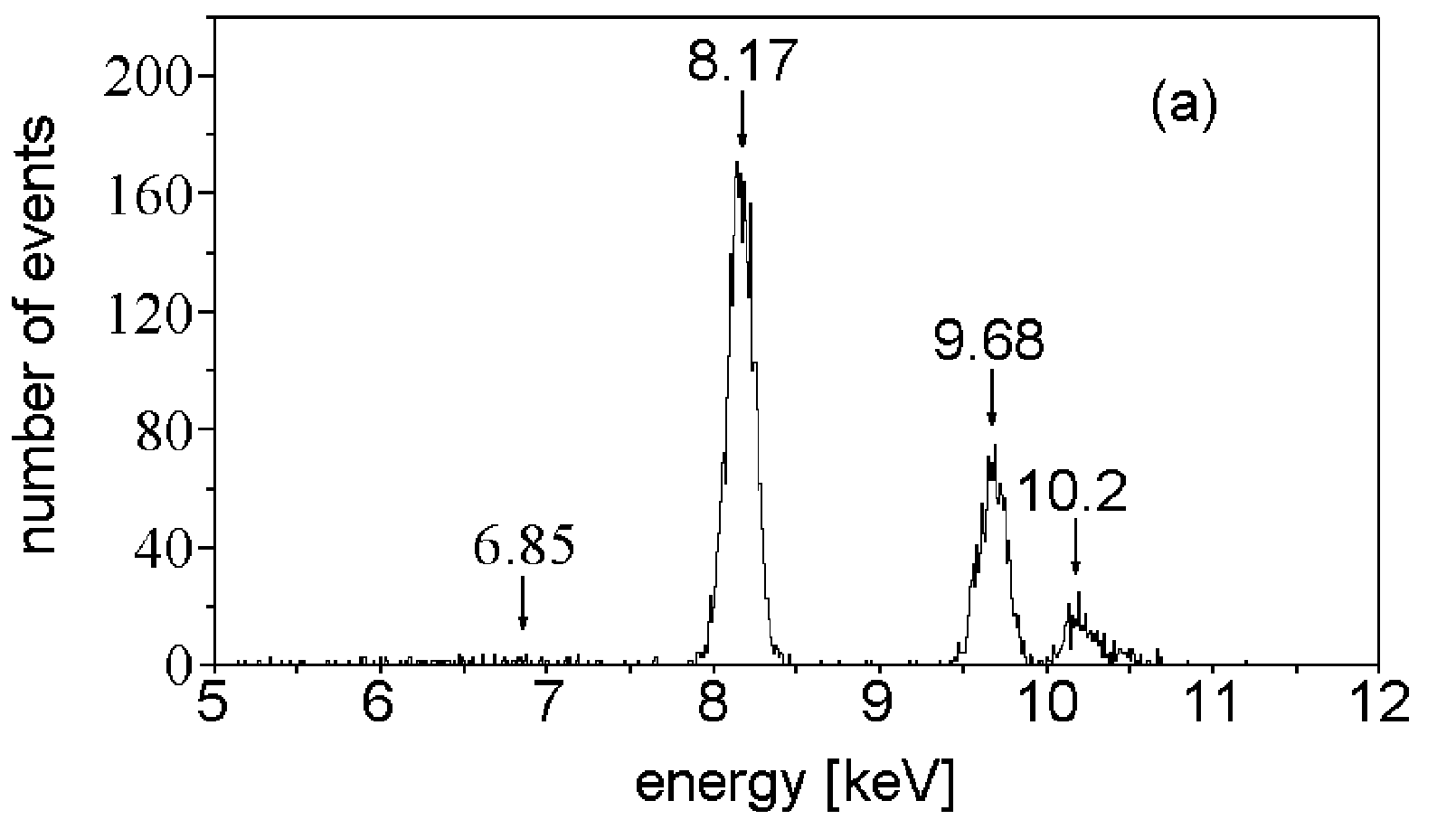}
\includegraphics[width=5.cm,angle=90]{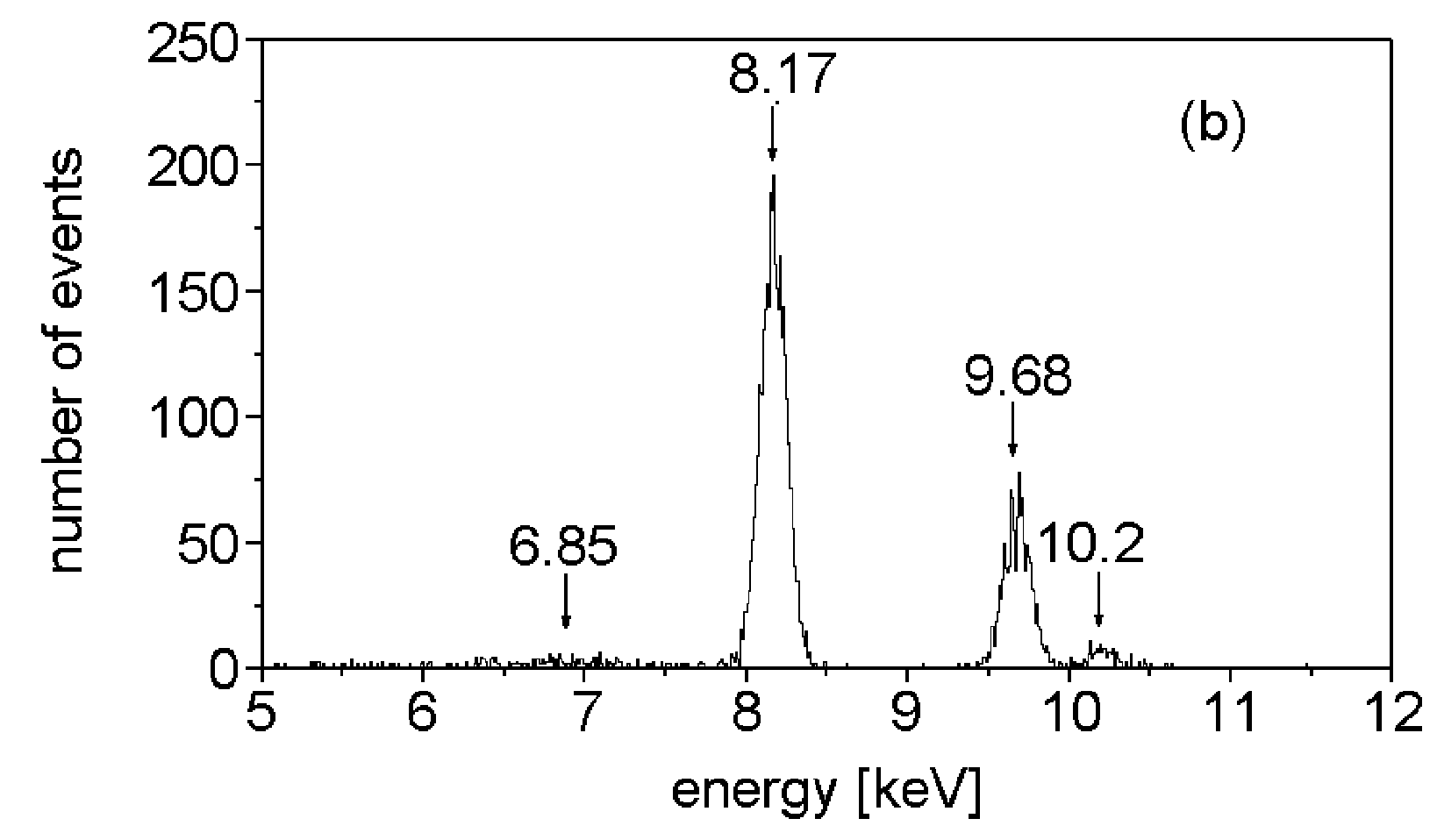}
                \caption{Energy spectra of the prompt event with
                        the del-$e$ criterion in runs III and IV\@.}
\label{fig:ene-pcd}
\end{figure}

\begin{figure}[ht]
\includegraphics[width=5.cm,angle=90]{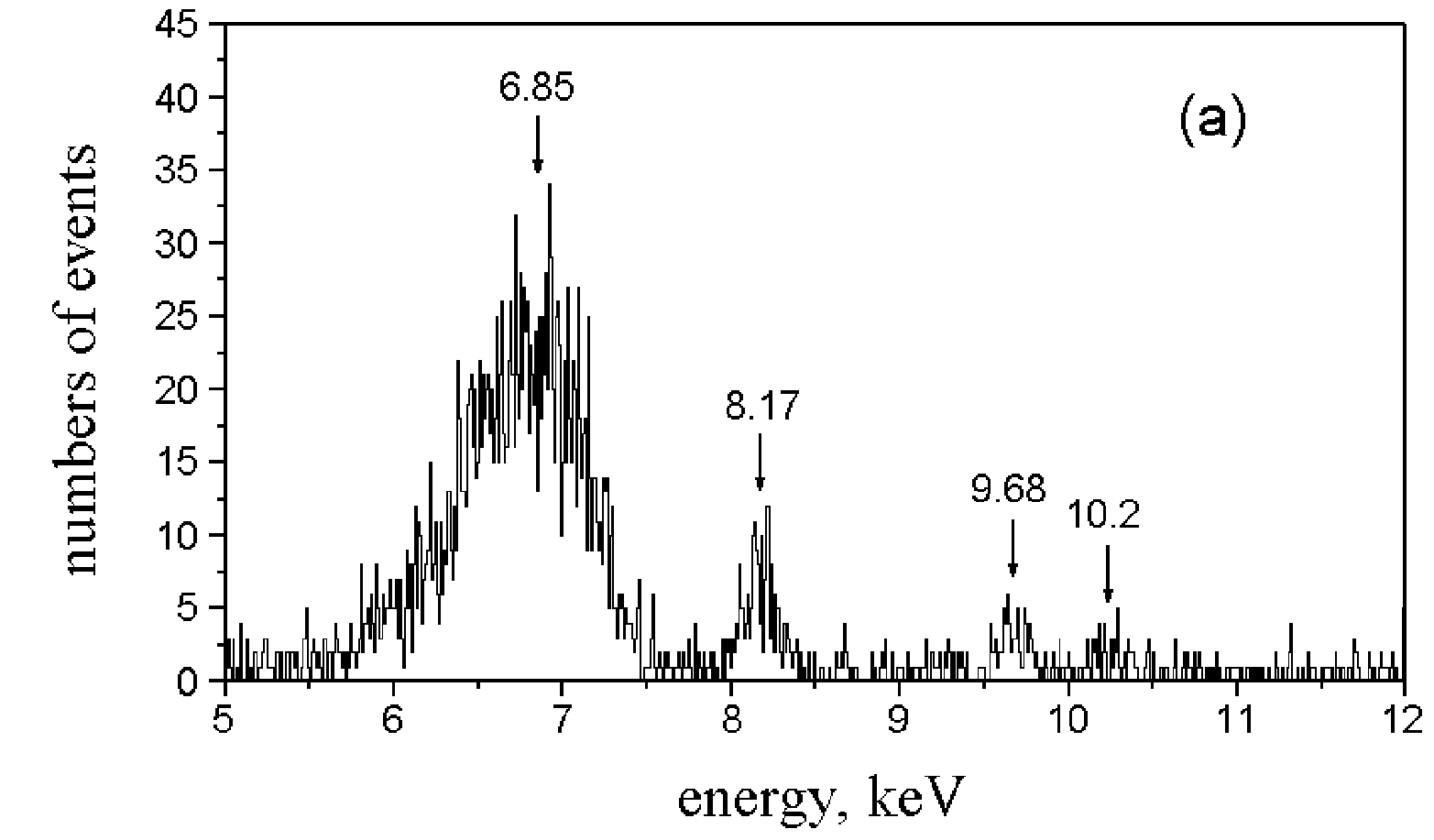}
\includegraphics[width=5.cm,angle=90]{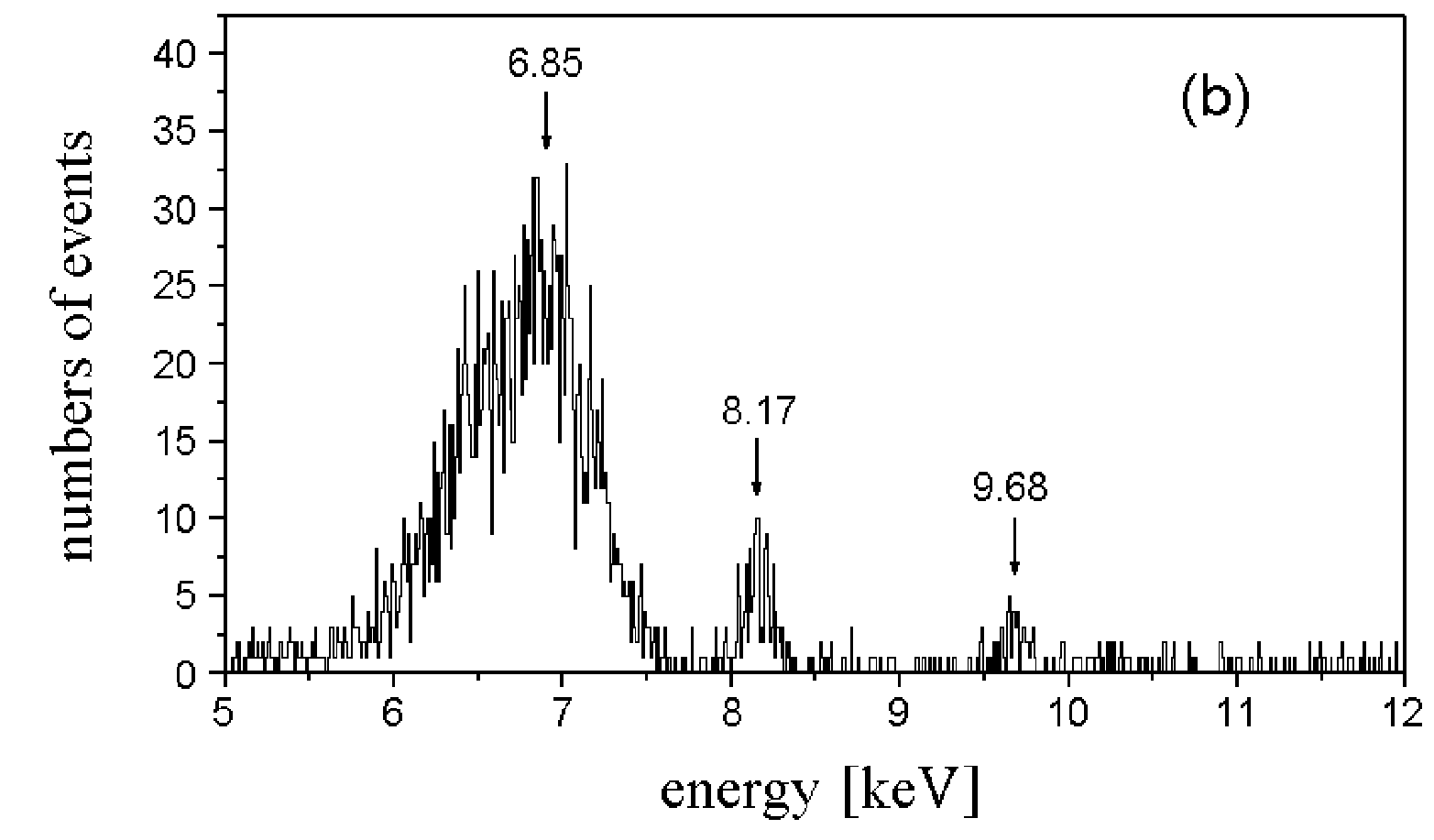}
                \caption{Energy spectra of the delayed event with the
                        del-$e$ criterion in runs III and IV\@.}
\label{fig:ene-dcd}
\end{figure}

\begin{figure}[ht]
\includegraphics[width=5.cm,angle=90]{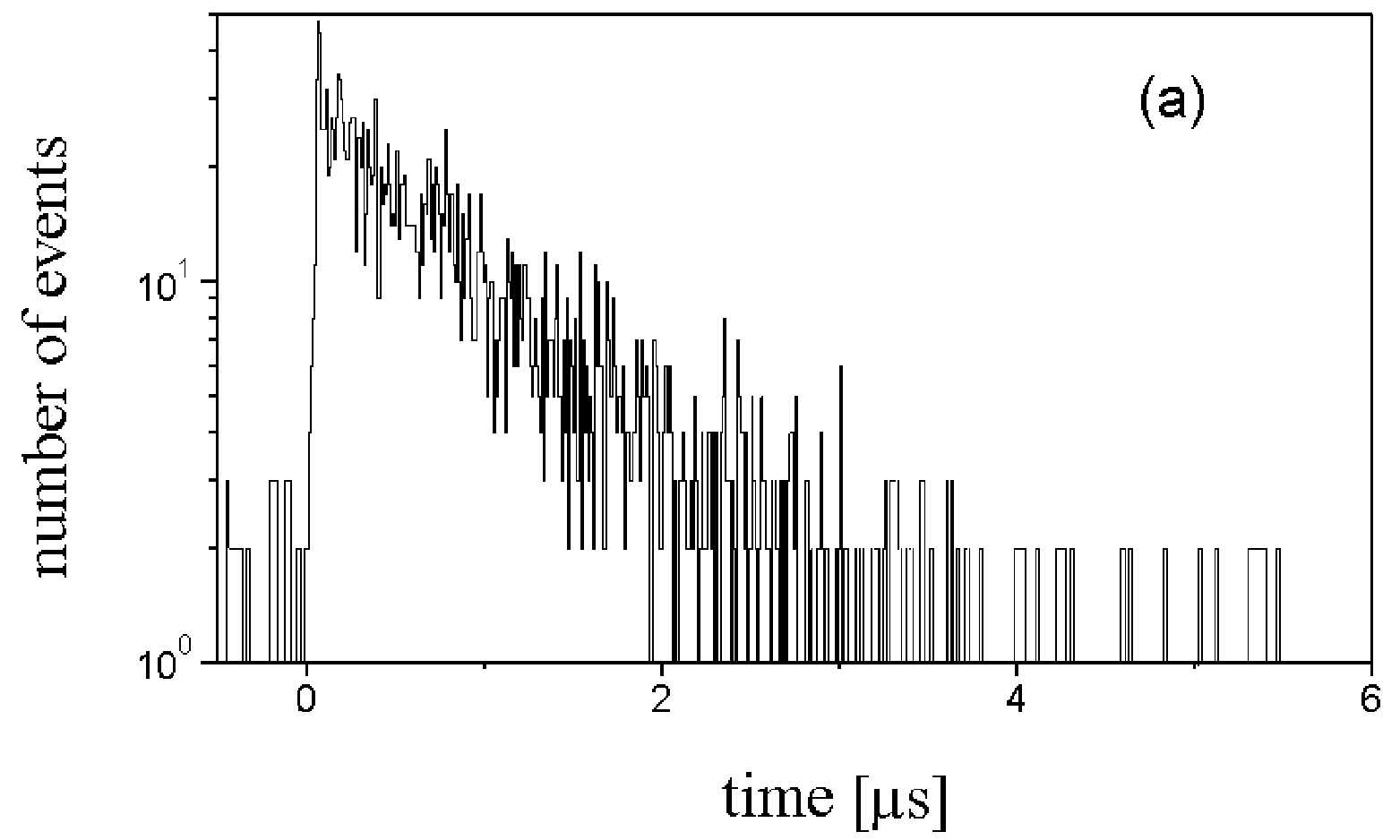}
\includegraphics[width=5.cm,angle=90]{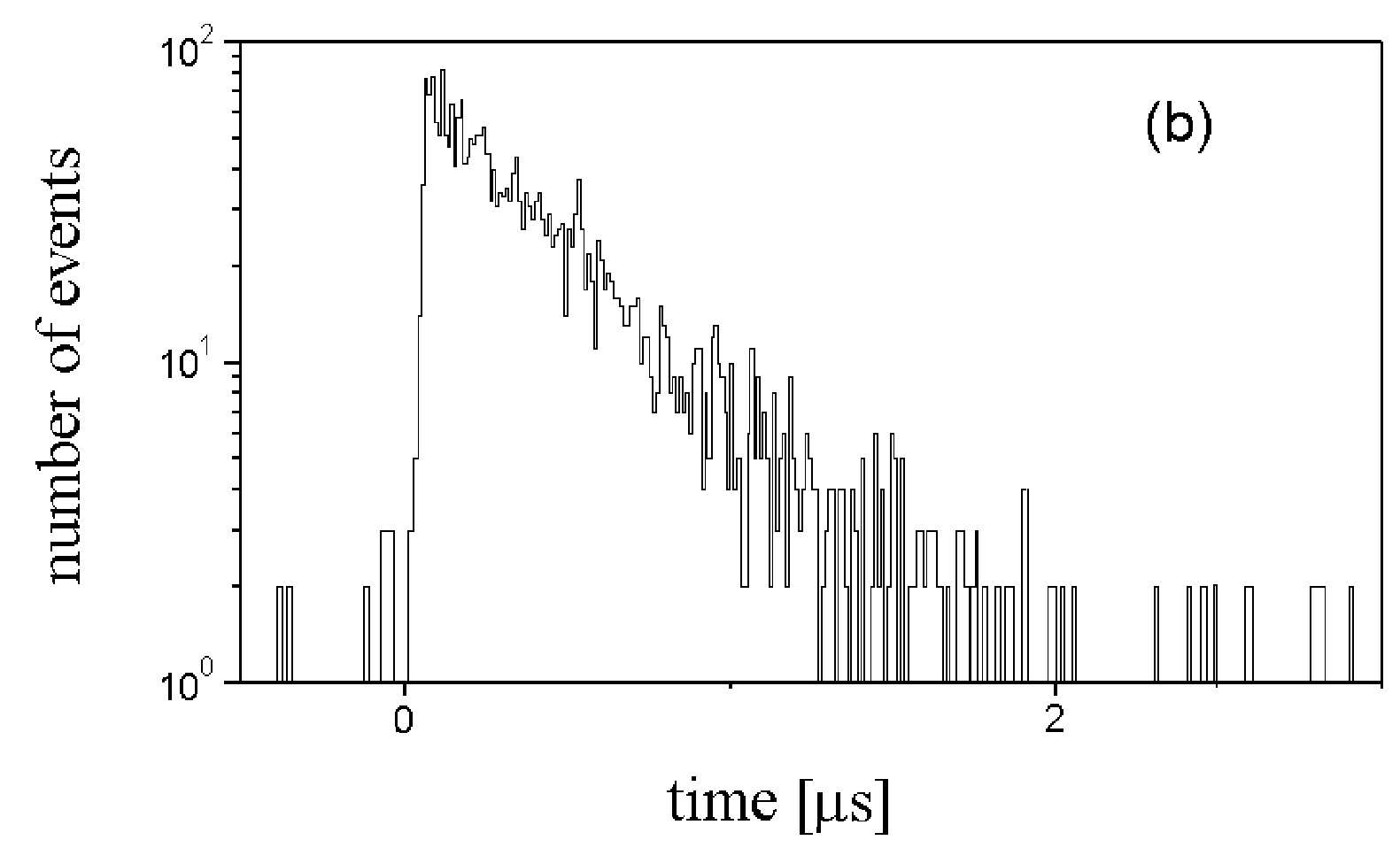}
                \caption{Delayed event time distributions with the
                        del-$e$ criterion in runs III and IV within
                        the energy range $[5.74-7.50]$~keV\@.}
\label{fig:time-cd}
\end{figure}

The experimental method to determine the $d\mu {}^{3}\mathrm{He}$
radiative decay branching ratio $\kappa_{d\mu\mathrm{He}}$ is
described in Sec.~\ref{su:k}.
Energy and time distributions of prompt and delayed events detected in
runs III and IV with muon decay electron coincidences are presented in
Figs.~\ref{fig:ene-pcd} to~\ref{fig:time-cd}.

To determine the $\lambda_{d\mu}$ and $\lambda_{d^3{\mathrm{He}}}$
rates (see Eq.~(\ref{eq29})) the $\gamma$--ray time distributions were
fitted within the energy range $[5.74-7.50]$~keV using the expression
\begin{eqnarray}
  \label{eq43}
  \frac{dN_{6.85}}{dt} & = & A^\gamma_{d\mu} \cdot e^{-\lambda_{d\mu}
        t} + A^\gamma_{\mathrm{Au}} \cdot e^{-\lambda_{\mathrm{Au}}t}
        + A^\gamma_{\mathrm{Al}} \cdot e^{-\lambda_{\mathrm{Al}}t}
        \nonumber \\
  & + & D^\gamma \cdot e^{-\lambda_0 t} + F^\gamma \, ,
\end{eqnarray}
where $A^\gamma_{d\mu}$, $A^\gamma_{\mathrm{Au}}$, and
$A^\gamma_{\mathrm{Al}}$ are the normalization constants of the
different target elements.
$D^\gamma$ and $F^\gamma$ are the constants describing the germanium
background.

The results of runs III and IV for the muonic deuterium ground state
disappearance rate and the molecular formation rate
$\lambda_{d^3{\mathrm{He}}}$ , using Eq.~(\ref{eq29}), are shown in
Table~\ref{tab:lambda}.
The averaged value $\lambda_{d^3{\mathrm{He}}} = 242 (20)\
\mu\mathrm{s}^{-1}$, where the errors include statistical as well as
systematic errors is consistent with the measurement of Maev
\textit{et al.}~\cite{maevx99b}, but is in disagreement with the work
of Gartner \textit{et al.}~\cite{gartn00}.

\begin{table}[t]
\begin{ruledtabular}
       \caption{Experimental results for the muonic deuterium ground
       state disappearance rate and the $d\mu {}^{3}\mathrm{He}$
       molecular formation rate.}
\label{tab:lambda}
\begin{tabular}{ccccccc}
Runs & $\lambda_{d\mu}$ & $\lambda_{d^3 {\mathrm{He}}}$ \\
    & $[\mu\mathrm{s}^{-1}]$ & $[\mu\mathrm{s}^{-1}]$  \\ \hline
III & $1.152 (36)_{stat} (30)_{syst}$  & $240 (13)_{stat} (15)_{syst}$ \\
IV  & $2.496 (58)_{stat} (100)_{syst}$ & $244 (6)_{stat} (16)_{syst}$
\\ \hline
Average &                                 & $242 (20)$ \\ \hline
Maev \textit {et al.}~\cite{maevx99b} & & $232 (9)$, $233
(16)$\footnotemark[1] \\
Gartner \textit {et al.}~\cite{gartn00}& & $185.6 (77)$ \\
\end{tabular}
            \footnotetext[1]{at 50~K and 39.5~K, respectively}
\end{ruledtabular}
\end{table}

According to Eq.~(\ref{eq31}) the determination of the branching ratio
$\kappa_{d\mu\mathrm{He}}$ requires the knowledge of both the total
number of $d\mu {}^{3}\mathrm{He}$ molecules formed in a mixture and
the number of $d\mu {}^{3}\mathrm{He}$'s decaying via the radiative
channel, Eq.~(\ref{eq1}a).
The corresponding numbers $N^{d\mu^3{\mathrm{He}}}_{tot}$ and
$N^{d\mu^3{\mathrm{He}}}_\gamma$ were determined using
Eqs.~(\ref{eq32}) and~(\ref{eq34}).
The $\gamma$~rays were measured during a time $t_\gamma$ and the
del-$e$ time interval was $t_e - t_\gamma$.
The $\varepsilon_{6.85}$ detection efficiency was determined using
detection efficiencies of $\mu{}^3\mathrm{He}$ atom $K$~series
transitions in runs I and II by a MC simulation.
This MC calculation took into account the $\eta_{6.85}$ attenuation of
$\gamma$~rays passing through all layers between the germanium
detector and the gas.
The time factors $f_t$ for the electrons and $F_t$ for the
$\gamma$~rays are slightly different for both runs, $f_t=0.84$ and
$F_t=0.94$ for run III, and $f_t=0.86$ and $F_t=0.99$ for run IV\@.
All results are presented in Table~\ref{tab:expdata}.

\begin{table*}[ht]
\begin{ruledtabular}
      \caption{Experimental results concerning formation and decay
              processes of $d\mu{}^{3}{\mathrm{He}}$ molecules
              obtained from runs III and IV\@.  ``Full'' stands for
              the full statistics, whereas del-$e$ represents the
              delayed electron criterion. The 6.85~keV $\gamma$~rays
              were measured within an energy range
              $[5.74-7.55]$~keV\@. The time intervals for the
              $\gamma$~rays and electrons are also given.}
\label{tab:expdata}
\begin{tabular}{cccccc}
Parameter & Units & \multicolumn{2}{c}{Run III} &
\multicolumn{2}{c}{Run IV} \\
& & full & del-$e$ & full & del-$e$ \\\hline
$t_\gamma$ & [$\mu$s] & $[(-0.03) - (+2.5)]$ & $[(-0.03) - (+2.5)]$ &
$[(-0.03) -  (+1.8)]$ & $[(-0.03) - (+1.8)]$ \\
$t_e - t_\gamma$ & [$\mu$s] & -- & $[0.08- 4.6]$ & -- & $[0.08 - 4.9]$ \\
$N_{6.85}$ &$[10^{3}]$& 17.42(21) & 2.15(6) & 20.07(23)&2.63(7) \\
$N^{d\mu^3{\mathrm{He}}}_{tot}$ &$[10^{3}]$& 20.81(136) &20.86(136) &16.50(70)
&16.41(72)\\
$N^{d\mu^3{\mathrm{He}}}_\gamma$&$[10^{3}]$& 4.20(10) &4.37(17) &3.76(10)
&3.53(18)\\
$\varepsilon_{6.85}(1-\eta_{6.85})$ & $[10^{-5}]$& 4.15(8) &5.76(15) &
6.26(19) & 8.72(32) \\
$\kappa_{d\mu\mathrm{He}}$ & & 0.203(14) & 0.209(17) & 0.228(12) &
0.213(15) \\
\end{tabular}
\end{ruledtabular}
\end{table*}

The $\kappa_{d\mu\mathrm{He}}$ values obtained in the present
experiment for two different $\mathrm{D}_2 +{}^{3}\mathrm{He}$
densities differ somewhat from the experimental result of
Ref.~\cite{augsb03}, i.e., $\kappa_{d\mu\mathrm{He}} = (0.301\pm
0.061)$ performed under slightly different experimental conditions
$(\varphi=0.697,\ c_{\mathrm{He}}=0.0913)$.
Our results differ slightly from the calculated
$\kappa_{d\mu\mathrm{He}}$ value in Ref.~\cite{czapl97b} for a total
angular momentum $J=0$ of the $d\mu {}^{3}\mathrm{He}$ complex.
However, they are in a good agreement with the calculations of
Refs.~\cite{kinox93,belya96} for a total angular momentum $J=1$.

A close comparison of the existing theoretical results for
$\kappa_{d\mu\mathrm{He}}$,
\cite{kinox93,gersh93,kravt93,czapl97b,belya95c,belya96}, with their
experimental results obtained in the present paper and in
Ref.~\cite{augsb03} may throw some light on the mechanism of
rotational $J=1 \to J=0$ transitions of $d\mu {}^{3}\mathrm{He}$
molecules in the $2p\sigma$ state, labeled $\tilde{\lambda}_{10}$ in
Fig.~\ref{fig:2}.
Specifically, two different mechanism of the $J=1 \to J=0$ transition
were proposed in Refs.~\cite{czapl96c,czapl02,czapl02a}
and~\cite{bogda98}.
Both mechanisms start with an Auger transition in a $d\mu +
{}^{3}\mathrm{He}$ collision,
\begin{equation}
  \label{eq44a}
  d \mu + {}^3\mathrm{He} \to \left
  [(d \mu {}^3\mathrm{He})^{++}_{2p\sigma,J=1} e \right ]^+ + e \, .
\end{equation}
The first mechanism~\cite{czapl96c,czapl02,czapl02a} consists of a two
stage process, namely the formation of a neutral complex in the
collision
\begin{eqnarray}
  \label{eq44}
  \left [ (d \mu{}^3\mathrm{He})^{++}_{2p\sigma,J=1} e \right ]^+ &+ &
  {\mathrm{He}} \qquad \stackrel{\lambda_{n}} \longrightarrow
  \nonumber \\
  \left [ (d \mu {}^3\mathrm{He})^{++}_{2p\sigma,J=1}2 e \right ] &+ &
  {\mathrm{He}}^+ \, ,
\end{eqnarray}
followed by a subsequent deexcitation due to external Auger effect
\begin{eqnarray}
  \label{eq45}
  \left [ (d \mu {}^3 \mathrm{He})^{++}_{2p\sigma,J=1}2 e \right ] & + &
  \mathrm{D}(\mathrm{D}_2) \qquad \stackrel{\lambda^{ext}_{Aug}}
  \longrightarrow \nonumber \\
  \left [ (d \mu {}^3\mathrm{He})^{++}_{2p\sigma,J=0}2 e \right ] & + &
  \mathrm{D}^+(\mathrm{D}^+_2) + e \, .
\end{eqnarray}

In the second mechanism~\cite{bogda98}, the $J=1 \to J=0$ transition
involves a number of molecular processes.
However, the corresponding transition rate is essentially determined
by a molecular cluster formation
\begin{eqnarray}
  \label{eq46}
  \left [ (d \mu {}^3\mathrm{He})^{++}_{2p\sigma,J=1} e \right ]^+ & + &
  \mathrm{D}_2 \qquad \stackrel{\lambda_{cl}} \longrightarrow
  \nonumber \\
  \left [ (d \mu {}^3\mathrm{He})^{++}_{2p\sigma,J=0} e \right
  ]\mathrm{D}_2 & &
\end{eqnarray}
and a subsequent inner electron conversion 
\begin{eqnarray}
  \label{eq47}
  \left [ (d \mu {}^3\mathrm{He})^{++}_{2p\sigma,J=1} e \right
  ]\mathrm{D}_2 & & \qquad \stackrel{\lambda^{int}_{Aug}}
  \longrightarrow \nonumber \\
  \left [ (d \mu {}^3\mathrm{He})^{++}_{2p\sigma,J=0} e \right
  ]\mathrm{D}_2^+ & + & e \, .
\end{eqnarray}

The first mechanism yields an effective $J=1 \to J=0$ transition rate
\begin{equation}
  \label{eq:47a}
  \tilde{\lambda}_{10} = \frac{\lambda_n\lambda^{ext}_{Aug}\varphi^2
  c_\mathrm{D} c_{\mathrm{He}}} {\lambda^1_{dec} +
  \lambda^{ext}_{Aug}\varphi c_\mathrm{D} + \lambda_n\varphi
  c_{\mathrm{He}}}\, ,
\end{equation}
the second mechanism gives
\begin{equation}
  \label{eq:47b}
  \tilde{\lambda}_{10} =
  \frac{\lambda_{cl}\lambda^{int}_{Aug}\varphi^2 c_\mathrm{D}}
  {\lambda^1_{dec} + \lambda^{int}_{Aug} + \lambda_{cl}\varphi
  c_\mathrm{D}}
\end{equation}
(see Refs.~\cite{bystr99c,bystr99d}).
The effective $d\mu {}^{3}\mathrm{He}$ decay rates are defined as
\begin{equation}
  \label{eq:47c}
  \lambda^J_{dec} = \lambda^J_\gamma + \lambda^J_e + \lambda^J_p \, ,
\end{equation}
for both rotational states, $J=0$ and $J=1$.

\begin{figure}[b]
\includegraphics[width=5.cm,angle=90]{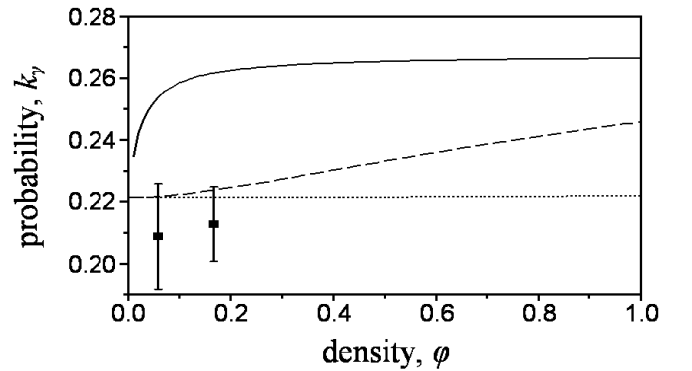}
                \caption{Density dependence of the $\gamma$--decay
                        branching ratio $\kappa_{d\mu\mathrm{He}}$.
                        Points with error bars are our experimental
                        values.  The solid line corresponds to the
                        second mechanism with
                        $\lambda_\mathrm{Aug}^{int} = 10^{12} \,
                        \mbox{s}^{-1}$~\cite{bogda98}. The dashed
                        lines represents the first mechanism with
                        $\lambda_\mathrm{Aug}^{ext} = 8.5 \times
                        10^{11} \, \mbox{s}^{-1}$~\cite{czapl96c},
                        whereas the dotted lines is given for
                        $\lambda_\mathrm{Aug}^{ext} = 10^{10} \,
                        \mbox{s}^{-1}$~\cite{czapl02,czapl02a}.}
\label{fig:density}
\end{figure}

Because the effective transition rate $\tilde{\lambda}_{10}$ is model
dependent, the ratio $\tilde{\lambda}_{10}/\lambda^1_{dec}$ may allow
us to check the validity of both models.
A proposal for a corresponding experiment was presented in
Refs.~\cite{bystr99c,bystr99d}.
It exploits the $J$--dependence of the probability for the radiative
$d\mu {}^{3}\mathrm{He}$ decay ratio $\kappa_{d\mu\mathrm{He}}$.
An unequivocal identification of the $J=1 \to J=0$ transition
mechanism should be possible by measuring the 6.85~keV $\gamma$--ray
yields for a series of different densities of $\mathrm{D}_2
+{}^{3}\mathrm{He}$ mixtures.
The density dependence of $\kappa_{d\mu\mathrm{He}}$ normalized to a
single $d\mu {}^{3}\mathrm{He}$ molecule is
\begin{equation}
  \label{eq:47d}
  \kappa_{d\mu\mathrm{He}} =
  \frac{1}{\lambda^1_{dec}+\tilde{\lambda}_{10}} \left [
  \lambda_\gamma^1 + \frac{\tilde{\lambda }_{10}\lambda^0_\gamma}
  {\lambda^0_{dec}}\right ]\, .
\end{equation}
Here, the decay rates $\lambda_{dec}^0=6 \times 10^{11}\
\mathrm{s}^{-1},\ \lambda_\gamma^0=1.8 \times 10^{11}\
\mathrm{s}^{-1}$~\cite{czapl97b}, $\lambda_{dec}^1=7 \times 10^{11}\
\mathrm{s}^{-1}$, and $\lambda_\gamma^1=1.55 \times 10^{11}\
\mathrm{s}^{-1}$ (obtained by averaging the corresponding results
taken from
Refs.~\cite{kinox93,gersh93,kravt93,korob93b,belya97,czapl97b,ishid93,
belya95c,belya96}) are model independent.
Concerning the first mechanism, we used $\lambda_n=2 \times 10^{13}\
\mathrm{s}^{-1},\ \lambda_{Aug}^{ext}=8.5 \times 10^{11}\
\mathrm{s}^{-1}$~\cite{czapl96c}, and $\lambda_{Aug}^{ext}=10^{10}\
\mathrm{s}^{-1}$~\cite{czapl02,czapl02a}.
For the second mechanism, we used $\lambda_{cl}=3 \times 10^{13}\
\mathrm{s}^{-1}$ and $\lambda_{Aug}^{int}=10^{12}\
\mathrm{s}^{-1}$~\cite{bogda98}.
All density dependent rates are normalized to LHD\@.

As can be seen from Fig.~\ref{fig:density}, our experimental values of
$\kappa_{d\mu\mathrm{He}}$ are in better agreement with the
theoretical results corresponding to the first mechanism as described
in Czapli\'nski \textit {et al.}~\cite{czapl96c,czapl02,czapl02a}.
More refined calculations of the $J=1 \to J=0$ transition including
realistic $(\mathrm{D}-d\mu {}^{3}\mathrm{He})^{0,(+\ \mathrm{or}\
2+)}$ interaction potentials have however to be performed before
definite conclusions can be drawn.
Calculations in Refs.~\cite{czapl02,czapl02a} go in this sense but
withing the framework of a semi--classical treatment.
Such a treatment seems rather problematic considering the collision
energies in such a system.
More accurate, i.e., purely quantum calculations are now in progress.


\subsection{Delayed $K$ series transitions of $\mu\mathrm{He}$  atoms}
\label{sec:delayed-k-series-1}

\begin{table}[b]
\begin{ruledtabular}
      \caption{Delayed muonic x--ray relative intensities for
              ${}^{3}\mathrm{He}$ and ${}^{4}\mathrm{He}$ atoms.}
\label{tab:delayed}
\begin{tabular}{cccc}
Run & Units & III $({}^{3}\mathrm{He})$ & IV $({}^{4}\mathrm{He})$  \\ \hline
$t_\gamma$ & [$\mu$s]& $[0.1 - 2.5]$ & $[0.1 - 1.8]$ \\
$t_e - t_\gamma$ & [$\mu$s] & $[0.08 - 4.6]$ & $[0.08 - 4.9]$ \\
$I_{del,\alpha}$ & [\%] & 0.605(75) &  0.728(85) \\
$I_{del,\beta}$  & [\%] & 0.185(47) &  0.160(48) \\
$I_{del,\gamma}$ & [\%] & 0.209(62) &  0.112(60) \\
\end{tabular}
\end{ruledtabular}
\end{table}

The relative intensities $I_{del,x}$ and $I_{del,x-\mathrm{e}}$ of
delayed $\mu\mathrm{He}$ $K$ series transitions were determined by
measuring the $N_{del,x}$ events during a time interval $ t_\gamma$
after the muon stop (see Table~\ref{tab:delayed}).
The corresponding relative intensities were obtained from the ratios
\begin{equation}
  \label{eq48}
  I_{del,x} =
  \frac{N_{del,x}}{\left[(1-\eta_x)\varepsilon_{x\alpha}\right]} \left
  / {\sum \limits_{x= \alpha, \beta, \gamma}
  \frac{N_{del,x}}{\left[(1-\eta_x)\varepsilon_{x\alpha}\right]}}
  \right . \, .
\end{equation}

Our results should, in principle, coincide with the prompt intensities
of $K$ series transitions if we assume that the incoming muon energy
distribution as well as the primary $\mu\mathrm{He}$ atom excited
states distribution due to direct muon capture are the same as the
corresponding ones for muons freed after the $d-d$ fusion.
The observed prompt relative intensities of the corresponding $K$
series transitions (see Table~\ref{tab:intens}) are however somewhat
different from the delayed ones indicating that the above conditions
are probably not fulfilled.


\section{Conclusions}
\label{sec:conclusions}

The measured relative intensities of $Kx$~line muonic x~rays in
$\mu{}^3\mathrm{He}$ and $\mu{}^{4}\mathrm{He}$ atoms (see
Tables~\ref{tab:intens} and~\ref{tab:intens-mix}) agree very well with
other experiments.
Only slight variation due either to the isotope or to the pressure are
visible.
The stopping power ratio $A$ of helium to deuterium atoms $A = 1.67
\left [ ^{+0.35}_{-0.33} \right]$ is also in good agreement with
earlier work~\cite{kottm93,bystr93d}.

Regarding the $\mathrm{q}_{1s}^{\mathrm{He}}$ probability for a $d\mu$
atom to reach its ground state in a $\mathrm{D}_2 +{}^{3}\mathrm{He}$
mixture at two different densities, our results are
\begin{eqnarray}
  \label{eq:49}
  \mathrm{q}_{1s}^{\mathrm{He}} = (0.882\pm 0.018) & & \varphi =0.0585
  \nonumber \\
  \mathrm{q}_{1s}^{\mathrm{He}} = (0.844\pm 0.020) & & \varphi = 0.1680
\end{eqnarray}
in agreement with theoretical calculations for an average $d\mu -
\mathrm{He}$ collision energy around 8~eV\@.

As for the $d\mu {}^{3}\mathrm{He}$ molecular formation rate
$\lambda_{d^3{\mathrm{He}}}$ for both our mixtures, our averaged value
is
\begin{equation}
  \label{eq:51}
  \lambda_{d^3{\mathrm{He}}} = (242 \pm 20) \ \mu\mathrm{s}^{-1} \, .
\end{equation}
Our result agrees very well with the measurement of Maev \textit{et
al.}~\cite{maevx99b}, but is in disagreement with the work of Gartner
\textit{et al.}~\cite{gartn00}.
This difference has not yet been understood.

Concerning the radiative decay branching ratio
$\kappa_{d\mu\mathrm{He}}$ for $d\mu {}^{3}\mathrm{He}$, also measured
for two different densities of the $\mathrm{D}_{2} + ^{3}\mathrm{He}$
mixture, the measured values,
\begin{eqnarray}
  \label{eq:50}
  \kappa_{d\mu\mathrm{He}} =  (0.203 \pm 0.014) & & \varphi =0.0585
  \nonumber \\
  \kappa_{d\mu\mathrm{He}} =  (0.228 \pm 0.012) & & \varphi = 0.1680 
\end{eqnarray}
are the same for both densities, but disagree with the recent results
by Augsburger \textit{et al.}~\cite{augsb03},
$\kappa_{d\mu\mathrm{He}} = (0.301 \pm 0.061)$, measured at a density
approximately two times bigger, namely $c_{\mathrm{He}} = 0.0913$.

Finally, the relative intensities of the delayed $K$ series
transitions $I^{\mathrm{D/He}}_{del,x}$ of $\mu\mathrm{He}$ atoms, due
to direct ${}^{3}\mathrm{He}$ muon capture or due to muon transfer
from deuterium to helium, after the muons were freed after $d-d$
fusion were also measured.
They differ from the prompt relative intensities, probably due to a
different primary distribution of excited states.

In conclusion, we were able to measure various interesting
characteristics of muon atom (MA) and muonic molecule (MM) processes
occurring in pure helium and in $\mathrm{D}_{2} + ^{3}\mathrm{He}$
mixtures with good accuracy.
This was possible by exploiting different germanium detectors for
$\gamma$--ray detection in a wide energy range $[3 \, \mbox{keV} - 10 \,
\mbox{MeV}]$, silicon Si(d$E-E$) telescope for the detection of
charged particles coming from nuclear fusion or nuclear muon capture
on ${}^{3}\mathrm{He}$ and muon decay electron detectors.
The self consistent methods increased the reliability of the presented
results.
Further measurements of quantities such as the muon stopping ratio
$A$, the $\mathrm{q}_{1s}^{\mathrm{He}}$ probability, and the
$\kappa_{d\mu\mathrm{He}}$ branching ratio in wider range of target
densities and helium concentrations should significantly improve the
accuracy of the corresponding values and clarify the complicated
picture of muonic processes occurring in deuterium--helium targets.


\begin{acknowledgments}

This work was supported by the Russian Foundation for Basic Research,
Grant No.~01--02--16483, the Polish State Committee for Scientific
Research, the Swiss National Science Foundation, and the Paul Scherrer
Institute.

\end{acknowledgments}


\end{document}